\documentclass[conference,a4paper]{IEEEtran}
\usepackage{xcolor}
\usepackage{balance}

\ifCLASSINFOpdf
  \usepackage[pdftex]{graphicx}
\else
  \usepackage[dvips]{graphicx}
\fi
\usepackage{amsmath,amssymb}
\ifCLASSOPTIONcompsoc
 \usepackage[caption=false,font=normalsize,labelfont=sf,textfont=sf]{subfig}
\else
 \usepackage[caption=false,font=footnotesize]{subfig}
\fi
\usepackage{cite}

\usepackage[colorlinks]{hyperref}
\usepackage{xcolor}
\hypersetup{
colorlinks,
    linkcolor={red!50!black},
    citecolor={blue!50!black},
    urlcolor={blue!80!black}
}

\usepackage{mathtools}

\def\XXint#1#2#3{{\setbox0=\hbox{$#1{#2#3}{\int}$}
     \vcenter{\hbox{$#2#3$}}\kern-.5\wd0}}

\usepackage{bm}% bold math
\newcommand{\R}{\mathbb{R}}
\newcommand{\C}{\mathbb{C}}

\newcommand{\ie}{\textit{i.e.}\/, }
\newcommand{\eg}{\textit{e.g.}\/, }
\newcommand{\cf}{\textit{cf.}\/, }

\providecommand*{\mrm}[1]{\mathrm{#1}}
\providecommand*{\unit}[1]{\ensuremath{\mrm{\,#1}}}
\providecommand*{\eu}{\ensuremath{\mrm{e}}}
\providecommand*{\iu}{\ensuremath{\mrm{i}}}

\renewcommand{\Re}{\ensuremath{\mrm{Re}}}	% The LaTeX standard is not ISO!
\renewcommand{\Im}{\ensuremath{\mrm{Im}}}	% The LaTeX standard is not ISO!

\providecommand*{\degree}{\ensuremath{^\circ}}

\begin{document}
%\title{A simple approach to line mixing combining narrowband resolution with far wing accuracy}
%\title{A modified projection approach to line mixing combining narrowband resolution with far wing accuracy}
\title{A modified projection approach to line mixing}

\author{\IEEEauthorblockN{
Sven~Nordebo\IEEEauthorrefmark{1},   % 1st author, 1st affiliations
%An~Author2\IEEEauthorrefmark{2},   % 2nd author, 2nd affiliations
%An~Author3\IEEEauthorrefmark{3}    % 3rd author, 3rd affiliations
%An~Author4\IEEEauthorrefmark{4}     % 4th author, 4th affiliations
}                                     % ...
%\\
\IEEEauthorblockA{\IEEEauthorrefmark{1}% 1st affiliations
Department of Physics and Electrical Engineering, Linn\ae us University,   351 95 V\"{a}xj\"{o}, Sweden. E-mail: sven.nordebo@lnu.se} 
% \\ E-mail: \{sven.nordebo,yevhen.ivanenko\}@lnu.se
%\IEEEauthorblockA{\IEEEauthorrefmark{2}
%Department of,.... E-mail: an.author2.@inst.se} 
%\IEEEauthorblockA{\IEEEauthorrefmark{3}
%Department of,.... E-mail: an.author3.@inst.se} 
%\IEEEauthorblockA{\IEEEauthorrefmark{4}
%Department of,.... E-mail: an.author4.@inst.se} 
}



\maketitle

\begin{abstract}
This paper presents a simple approach to combine the high-resolution narrowband features of some
desired isolated line models together with the far wing behavior of the projection based strong collision (SC) method to line mixing which was 
introduced by Bulanin, Dokuchaev, Tonkov and Filippov. 
The method can be viewed in terms of a small diagonal perturbation
of the SC relaxation matrix providing the required narrowband accuracy close to the line centers at the same time
as the SC line coupling transfer rates are retained and can be optimally scaled to thermalize the radiator after impact.
The method can conveniently be placed in the framework of the Boltzmann-Liouville transport equation where a rigorous diagonalization of the line mixing problem
requires that molecular phase and velocity changes are assumed to be uncorrelated.
A detailed analysis for the general Doppler case is given based on the first order Rosenkranz approximation,
and which also provides the possibility to incorporate quadratically speed dependent parameters.
Exact solutions for pure pressure broadening and explicit Rosenkranz approximations are given 
in the case with velocity independent parameters (line frequency, strength, width and shift) 
which can readily be retrieved from databases such as HITRAN for a large number of species.
Numerical examples including comparisons to published measured data are provided
in two specific cases concerning the absorption of carbon dioxide in its infrared band of asymmetric stretching, as well as 
of atmospheric water vapor and oxygen in relevant millimeter bands.
\end{abstract}

\vskip0.5\baselineskip
\begin{IEEEkeywords}
Wide-band molecular absorption spectra, collisional broadening, Doppler broadening, line mixing, spectral line shapes, radiative transfer in the atmosphere.
\end{IEEEkeywords}

\section{Introduction}

The presence of line mixing has been recognized as the main reason for the deviation of actual far wing line shapes 
from those that are calculated as a simple sum of Lorentzian lines, see \eg \cite[Fig.~1]{Filippov+etal2002}, \cite[Fig.~2]{Tonkov+Filippov2003} and \cite[Fig.~IV.7 on p.~188]{Hartmann+etal2021}. 
The effects of incomplete (soft) collisions does only play a minor role and which therefore motivates the use of the hard collision models 
within the impact approximation, see \eg \cite{Tonkov+Filippov2003,Hartmann+etal2021}. 
However, it is also commonly understood that a unified treatment of all parts of the spectrum through a relaxation matrix remains an open issue which requires 
the inclusion of an interaction potential taking all collisional coupled lines into account via frequency dependent relaxation coefficients, \cf \cite[p.~460]{Hartmann+etal2021}.

In this paper, we will take a more simplified and pragmatic viewpoint.
We will investigate the possibility of complementing the high-resolution narrowband features of some
existing isolated line models with the line coupling transfer rates that are obtained from the strong collision (SC) method introduced 
by Bulanin, Dokuchaev, Tonkov and Filippov in \eg
\cite{Dokuchaev+etal1982,Bulanin+etal1984,Filippov+Tonkov1993,Tonkov+etal1996,Filippov+etal2002,Tonkov+Filippov2003}, \cf also \cite[p.~211]{Hartmann+etal2021}. 
The aim is to derive a simple and flexible approach to line mixing with high accuracy close to the line centers as well as in the far wings.
For ease of implementation it is furthermore required that the spectral function can be implemented as a simple sum of individual lines based on a 
rigorous (within first order approximation) diagonalization of the line mixing problem.
The method should finally be able to take as sole input parameters the molecular transition frequencies, the line strengths and the line widths and shifts that are readily available for 
a large variety of species in spectroscopic databases such as \eg HITRAN
\cite{HITRANorg,HITRANdefs,Rothman+etal1998,Simeckova+etal2006,Gordon+etal2022}.
This will then provide a practical compromise towards a unified treatment for all species in all parts of the spectrum.

The proposed approach can be motivated by the computationally exhaustive line-by-line calculations of broadband radiative transfer in the atmosphere, 
see \eg \cite[p.~126]{Liou2002} and \cite{Berk+Hawes2017,Nordebo2021a}.
It is for this reason that simple isolated line shapes are usually employed for this purpose, but it is quite unclear how much the lack of far wing accuracy and consequently 
the overestimated atmospheric absorption may affect the results. This is not an issue in many remote sensing applications where it is only the small scale details of the spectrum that is in focus.
But the far wing behavior of a spectral band may be of major importance for providing the correct ``base-line'' level of atmospheric 
absorption in radiative transfer analysis. This may be crucial \eg in climate modeling to make the correct predictions concerning the greenhouse effect.
Incorrectly modelled Lorentzian line shapes in the far end of a ro-vibrational band may furthermore mask the true spectral signatures of the atmosphere
and thereby obscuring the possibility of exploring new potential applications in remote sensing and climate surveillance.

In this paper, we will adopt as a general framework the kinetic equation method by 
Rautian and Sobelman \cite{Rautian+Sobelman1967,Hartmann+etal2021,Tran+Hartmann2009,Ngo+etal2012,Ngo+etal2013,Tennyson+etal2014,Lance+etal1997,Pine1999}
which can readily be adapted to include the line mixing effects as described in \eg \cite{Smith+etal1971a,Smith+etal1971b,Ciurylo+Pine2000}.
There is a vast literature in this field, see \eg \cite{Hartmann+etal2021} with references, and
among the pioneering work is in particular worth mentioning Kolb and Griem \cite{Kolb+Griem1958}, Baranger \cite{Baranger1958}, Gordon \cite{Gordon1966,Gordon1967} 
and Rosenkranz \cite{Rosenkranz1975}. 
Later developments include the so called ``beyond Voigt'' profiles such as the Hartmann-Tran (HT) profile \cite{Ngo+etal2013,Pine1999} encompassing
partially correlated collisions and speed dependent pressure broadening and shifts, and
which is now becoming standardized in spectroscopic databases such as HITRAN \cite{Tennyson+etal2014,Gordon+etal2022}.
In particular, the papers \cite{Ngo+etal2012,Ngo+etal2013} provide efficient numerical procedures for evaluation of the physical models provided 
in \cite{Lance+etal1997,Pine1999} in case of quadratic speed dependence of collisional width and shift, \cf also \cite{Rohart+etal1994,Boone+etal2007}.
However, following \cite{Ciurylo+Pine2000}, it turns out that a rigorous diagonalization of the line mixing problem based on the kinetic equation method
can not generally be achieved with correlated collisions, 
line dependent Dicke narrowing and speed dependent pressure broadening and shifts,
see in particular \cite[Appendix~A]{Ciurylo+Pine2000}.
The more general line mixing methods may therefore be computationally huge, 
\cf \eg \cite[p.~379 and Eq.~(2.13)]{Ciurylo+Pine2000} and \cite[p.~73]{Pine+Gabard2000}.
Nevertheless, it has been reported that the isolated HT line shapes can take line mixing effects into account by simple empirical modifications, 
\cf \eg \cite[Eq.~(9)]{Ngo+etal2013}, \cite[Eq.~(15)]{Pine+Gabard2000} and \cite[Eq.~(6)]{Vasilchenko+etal2023},
see also \cite{Rohart+etal1994,Boone+etal2007,Pine+Markov2004,Pine+Gabard2003,Pine2019,Domyslawska+etal2022,Boulet+Hartmann2021} 
for further theory and applications in this regard.

It has been reported that the strong collision (SC) method by Filippov et al \cite{Dokuchaev+etal1982,Bulanin+etal1984,Filippov+Tonkov1993,Tonkov+etal1996,Filippov+etal2002,Tonkov+Filippov2003}
suffers from the major disadvantages that it does not take into account the distinguishing features of various perturbing gases nor of the different ro-vibrational branches 
of the radiator, see \eg \cite[p.~131]{Tonkov+Filippov2003}. In particular, the model actually produces the same line width for all lines 
and which may naturally induce very large errors close to the line centers.
However, as will be demonstrated in this paper, it will only require a small adjustment of this model based on a diagonal perturbation to obtain a line-mixing model that is able to take
existing line specific data of various host gases and radiators into account and thus providing a high accuracy close to the line centers as well as in the far wing. 
The latter is achieved by proper scaling of the remaining line coupling transfer rates of the SC relaxation matrix.
The so perturbed relaxation matrix will maintain the condition of detailed balancing,
but its property as a rigorous projector will be slightly relaxed in favor of the new desired features.
Notably, this projection property is equivalent to a particular sum rule which has been derived under the assumption that the absorbing molecule is a rigid rotor and that the line coupling
depend only on its rotational states, \cf \cite[Eq.~(A9)]{Bulanin+etal1984} and \cite[Eq.~(IV.14) on p.~193]{Hartmann+etal2021}. 
However, since the proposed modified projection method is taking all the available transitions into account and not only some specific ro-vibrational branches, 
this sum rule is no longer strictly required here.
It should therefore be physically sound to relax this sum rule in favor of more accurate line centers at the same time as the remaining line coupling transfer rates are 
optimally scaled in a way as to approximately thermalize (project) all the available states after impact. 
The precise meaning of this general description will be detailed in the sections that follow. 
Numerical examples with comparisons to published measured data will finally be included to demonstrate the usefulness of this approach.

The rest of the paper is organized as follows. In section \ref{sect:General} is formulated the general framework for line mixing setting the notation
for the modified projection approach which is detailed in section \ref{sect:ModProj}. The results are then summarized more conveniently in wavenumber domain in section \ref{sect:wavenumber} 
including a calculation of the absorption coefficient taking the fluctuation-dissipation theorem into account. The numerical examples are given in section \ref{sect:NumEx}, the summary in section \ref{sect:Summary}
and an appendix is finally included to review some important integrals associated with the Faddeeva function.

\section{A general framework for line mixing}\label{sect:General}
 
As a standard hard collision line mixing model we consider here the following Boltzmann-Liouville transport equation formulated in velocity space as
\begin{multline}\label{eq:dFdtpartcorr}
\frac{\partial}{\partial t}F_n(\bm{v},t)=-\iu\left(\omega_{0n} + \bm{k}\cdot\bm{v} \right) F_n(\bm{v},t)-\beta_n F_n(\bm{v},t) \\
+\beta_n f(\bm{v})\int_{\bm{v}^\prime}F_n(\bm{v}^\prime,t)\mrm{d}\bm{v}^\prime 
-\sum_{n^\prime=1}^N W_{nn^\prime}F_{n^\prime}(\bm{v},t) \\
-f(\bm{v})\sum_{n^\prime=1}^N C_{nn^\prime} \int_{\bm{v}^\prime} F_{n^\prime}(\bm{v}^\prime,t)\mrm{d}\bm{v}^\prime,
\end{multline}
\cf \cite{Smith+etal1971a,Smith+etal1971b,Ciurylo+Pine2000}, and where the notation has been largely adapted to \cite{Ciurylo+Pine2000}.
Here, $\bm{v}$ is the velocity, $\omega_{0n}$ is the transition frequency of a particular line $n$ and the factor $\bm{k}\cdot\bm{v}$ models the Doppler dephasing 
where $\bm{k}=k\hat{\bm{k}}$ is the wave vector, $k=\omega/\mrm{c}_0$ the wavenumber of the incident radiation and $\mrm{c}_0$ the speed of light in vacuum.
Further, $W_{nn^\prime}$ are the relaxation coefficients for the phase-changing collisions which are uncorrelated with velocity changes and 
$C_{nn^\prime}$ are the relaxation coefficients for the correlated phase-changing collisions which are simultaneously changing the velocity of the molecule.
The parameters $\beta_n$ are the line specific velocity-changing collision rates. 
The Maxwell-Boltzmann distribution is given by $f(\bm{v})=(\sqrt{\pi}\widetilde{v})^{-3}\eu^{-(v/\widetilde{v})^2}$
where $v=|\bm{v}|$, $\widetilde{v}=\sqrt{2k_\mrm{B}T/\mu_\mrm{r}}$ is the most probable speed, $k_\mrm{B}$ the Boltzmanns constant, 
$T$ the temperature and $\mu_\mrm{r}$ the mass of the radiator.
It is furthermore assumed here that the parameters $C_{nn^\prime}$ and $\beta_n$ are independent of velocity $\bm{v}$.
The off-diagonal elements $W_{nn^\prime}$ are the negative of the line coupling transfer rates and
the diagonal elements $W_{nn}=\gamma_{0n}(v)+\iu\delta_{0n}(v)$ consist of the broadening and frequency shift parameters
that generally may depend on speed. The parameters $W_{nn^\prime}$, $C_{nn^\prime}$ and $\beta_n$ are typically taken to be linear with pressure $p$.
It is noted here that even though our numerical examples below will deal solely with uncorrelated collisions, high-pressure (no Doppler shift), no Dicke narrowing
and velocity independent width and shift parameters, we will keep the formulation as general as possible in order to facilitate future developments 
employing \eg the HT-profile as a model for the required line centers.

The total dipole autocorrelation function is represented here in velocity space as $C(t)=\int_{\bm{v}}C(\bm{v},t)\mrm{d}\bm{v}$ where 
\begin{equation}\label{eq:Cvtsumdef}
C(\bm{v},t)=\sum_n \mu_n F_n(\bm{v},t),
\end{equation}
and where $\mu_n$ are the transition dipole moments associated with a particular line. 
Based on the dipole approximation ($\eu^{\iu \bm{k}\cdot\bm{r}}\approx 1$ where $\bm{r}$ is the position of charges)
the transition dipole moments can be assumed here to be real valued.
We have also
\begin{equation}\label{eq:Fntdef}
F_n(t)=\int_{\bm{v}}F_n(\bm{v},t)\mrm{d}\bm{v},
\end{equation}
so that the total dipole autocorrelation function is given by
\begin{equation}\label{eq:Ctsumdef}
C(t)=\sum_n \mu_n F_n(t).
\end{equation}
The initial condition in the hard collision model \eqref{eq:dFdtpartcorr} is based on an assumption of thermal equilibrium at time zero,
and is hence given by $F_n(\bm{v},0)=f(\bm{v})\rho_n\mu_n$ 
where $\rho_n$ is the canonical (thermal equilibrium) density associated with the 
lower transition level of the unperturbed absorbing molecule.
We have thus $F_n(0)=\rho_n\mu_n$ and we can now define the band strength as $C(0)=\sum_n S_n$ where $S_n=\rho_n\mu_n^2$ is the line strength.
Finally, the condition of detailed balancing is assumed for the relaxation matrix so that $W_{kn}\rho_n=\rho_kW_{nk}$, \cf \cite{Ciurylo+Pine2000}.

We denote by $\widetilde{C}(\omega)=\int_0^\infty C(t)\eu^{\iu\omega t}\mrm{d}t$ the Fourier-Laplace transform of the dipole autocorrelation function $C(t)$ having 
symmetry $C(-t)=C^*(t)$.
The spectral density is then given by $I(\omega)=\frac{1}{2\pi}\int_{-\infty}^{\infty}C(t)\eu^{\iu\omega t}\mrm{d}t=\frac{1}{\pi}\Re\{\widetilde{C}(\omega)\}$
for $\omega\in\R$ where we are assuming that the real line belongs to the region of convergence of the corresponding Fourier-Laplace transform.
Now, by using quantum mechanical principles, it can be shown that the absorption coefficient of a gaseous media, $\sigma_\mrm{a}$ (in \unit{m^2/molecule}), 
can be expressed quite generally in SI-units as\footnote{It is noticed that this formula is most oftenly referred to in Gaussian units as
$\sigma_\mrm{a}=\frac{4\pi^2\omega}{3\hbar\mrm{c_0}}I(\omega)\left(1-\eu^{-\beta\hbar\omega}\right)$, 
\cf \eg \cite[p.~3084]{Gordon1966a}, \cite[p.~2348]{Robert+Galatry1971}, \cite[p.~111]{Filippov+Tonkov1993} and \cite[p.~14]{Hartmann+etal2021}.}
\begin{equation}\label{eq:sigmaaexpr}
\sigma_\mrm{a}(\omega)=\frac{\pi\eta_0\omega}{3\hbar}(1-\eu^{-\beta\hbar\omega})I(\omega),
\end{equation}
where $I(\omega)$ is the spectral density defined as above, $\omega$ the angular frequency, $\eta_0$ the wave impedance of vacuum,
$\beta=1/k_\mrm{B}T$ and $\hbar=h/2\pi$ where $h$ is Plancks constant \cf \eg \cite[p.~14]{Hartmann+etal2021}.
Here, the factor $1-\eu^{-\beta\hbar\omega}$ above is due to the fluctuation-dissipation theorem stating that 
\begin{equation}
I(-\omega)=\eu^{-\beta\hbar\omega}I(\omega), 
\end{equation}
\cf \cite[p.~19]{Hartmann+etal2021}.
It is finally noticed here that the frequency dependent parameter $k=\omega/\mrm{c}_0$ relating to the wavenumber of the incident radiation 
is present already in the time-domain in the expression \eqref{eq:dFdtpartcorr} above.

\subsection{Solving the line mixing problem}
We will now formulate the solution to \eqref{eq:dFdtpartcorr} based on various simplifying assumptions.
In essence, this will follow the developments made in \cite{Ciurylo+Pine2000}.
To simplify the derivations we introduce a vector-matrix notation where ${\bf F}(\bm{v},t)$ and $\bm{\mu}$ are column vectors with elements $F_n(\bm{v},t)$ and $\mu_n$, 
${\bf W}$ and  ${\bf C}$ are matrices with elements $W_{nn^\prime}$ and $C_{nn^\prime}$, and $\bm{\beta}$, $\bm{\omega}_0$ and $\bm{\rho}$ are the diagonal matrices 
$\bm{\beta}=\mrm{diag}\{\beta_n\}$, $\bm{\omega}_0=\mrm{diag}\{\omega_{0n}\}$ and $\bm{\rho}=\mrm{diag}\{\rho_n\}$, respectively. 
The identity matrix is denoted ${\bf I}$.
The line mixing problem \eqref{eq:dFdtpartcorr} can now be formulated as the following initial value problem for $t\geq 0$
\begin{equation}\label{eq:dFdtLinemix}
\left\{\begin{array}{l}
\displaystyle\frac{\partial}{\partial t}{\bf F}(\bm{v},t)=-\left( {\bf W}+\bm{\beta}+\iu\left(\bm{\omega}_0+\bm{k}\cdot\bm{v}\cdot {\bf I}\right)\right) {\bf F}(\bm{v},t) \vspace{0.2cm} \\
\hspace{3.0cm} +f(\bm{v})\left( \bm{\beta}- {\bf C} \right)\int_{\bm{v}^\prime}{\bf F}(\bm{v}^\prime,t)\mrm{d}\bm{v}^\prime \vspace{0.2cm}  \\
{\bf F}(\bm{v},0) = f(\bm{v})\bm{\rho}\bm{\mu}.
\end{array}\right.
\end{equation}
We define also the vector ${\bf F}(t)$ with elements $F_n(t)$, so that the correlation function can be expressed as
$C(t)=\bm{\mu}^\mrm{T}{\bf F}(t)$ and its Fourier-Laplace transform as $\widetilde{C}(\omega)=\bm{\mu}^\mrm{T}\widetilde{\bf F}(\omega)$
where $\widetilde{\bf F}(\omega)$ is the Fourier-Laplace transform of the vector ${\bf F}(t)$.
From \eqref{eq:Fntdef} we have also that $\widetilde{\bf F}(\omega)=\int_{\bm{v}}\widetilde{\bf F}(\bm{v},\omega)\mrm{d}\bm{v}$
where $\widetilde{\bf F}(\bm{v},\omega)$ is the Fourier-Laplace transform of the vector ${\bf F}(\bm{v},t)$.

We proceed now by taking the Fourier-Laplace transform of \eqref{eq:dFdtLinemix} yielding
\begin{multline}
-\iu\omega \widetilde{\bf F}(\bm{v},\omega)-f(\bm{v})\bm{\rho}\bm{\mu} \\
=-\left( {\bf W}+\bm{\beta}+\iu\left(\bm{\omega}_0+\bm{k}\cdot\bm{v}\cdot {\bf I}\right)\right) \widetilde{\bf F}(\bm{v},\omega) \\
+f(\bm{v})\left( \bm{\beta}- {\bf C} \right)\int_{\bm{v}^\prime}\widetilde{\bf F}(\bm{v}^\prime,\omega)\mrm{d}\bm{v}^\prime, 
\end{multline}
or
\begin{multline}\label{eq:Ftildeeq1}
\left( {\bf W}+\bm{\beta}-\iu\left(\omega\cdot{\bf I}-\bm{\omega}_0-\bm{k}\cdot\bm{v}\cdot {\bf I}\right)\right) \widetilde{\bf F}(\bm{v},\omega) \\
-f(\bm{v})\left( \bm{\beta}- {\bf C} \right)\int_{\bm{v}^\prime}\widetilde{\bf F}(\bm{v}^\prime,\omega)\mrm{d}\bm{v}^\prime 
=f(\bm{v})\bm{\rho}\bm{\mu}. 
\end{multline}
Following the same procedure as in \cite{Ciurylo+Pine2000}, we introduce now
\begin{equation}\label{eq:Gvomegadef}
{\bf G}(\bm{v},\omega)=\left( {\bf W}+\bm{\beta}-\iu\left(\omega\cdot{\bf I}-\bm{\omega}_0-\bm{k}\cdot\bm{v}\cdot {\bf I}\right)\right)^{-1},
\end{equation}
so that
\begin{multline}\label{eq:Ftildeeq2}
 \widetilde{\bf F}(\bm{v},\omega) 
-f(\bm{v}){\bf G}(\bm{v},\omega)\left( \bm{\beta}- {\bf C} \right)\int_{\bm{v}^\prime}\widetilde{\bf F}(\bm{v}^\prime,\omega)\mrm{d}\bm{v}^\prime \\
=f(\bm{v}){\bf G}(\bm{v},\omega)\bm{\rho}\bm{\mu}.
\end{multline}
Next, by introducing
\begin{equation}\label{eq:Gomegadef}
{\bf G}(\omega)=\int_{\bm{v}}f(\bm{v}){\bf G}(\bm{v},\omega)\mrm{d}\bm{v},
\end{equation}
and integrating \eqref{eq:Ftildeeq2} over velocity space, we obtain
\begin{equation}\label{eq:Ftildeeq3}
 \widetilde{\bf F}(\omega) 
-{\bf G}(\omega)\left( \bm{\beta}- {\bf C} \right)\widetilde{\bf F}(\omega)
={\bf G}(\omega)\bm{\rho}\bm{\mu}.
\end{equation}
The solution to \eqref{eq:Ftildeeq3} can thus be expressed as
\begin{equation}
\widetilde{\bf F}(\omega)=\left({\bf I}- {\bf G}(\omega)\left( \bm{\beta}- {\bf C} \right) \right)^{-1}{\bf G}(\omega)\bm{\rho}\bm{\mu}.
\end{equation}
The Fourier-Laplace transform $\widetilde{C}(\omega)$ is now given by
\begin{multline}\label{eq:Ctildeomegasol}
\widetilde{C}(\omega)=\bm{\mu}^\mrm{T}\widetilde{\bf F}(\omega) \\
=\bm{\mu}^\mrm{T}\left({\bf I}- {\bf G}(\omega)\left( \bm{\beta}- {\bf C} \right) \right)^{-1}{\bf G}(\omega)\bm{\rho}\bm{\mu}.
\end{multline}
This is the result given in \cite[Eq.~(2.13)]{Ciurylo+Pine2000}.

In order to effectively exploit a diagonalization of the matrix ${\bf W}+\bm{\beta}+\iu\bm{\omega}_0$ and to write \eqref{eq:Ctildeomegasol}
as a sum over individual lines, we will see below that it will prove to be very convenient if
the eigenvectors of ${\bf W}+\iu\bm{\omega}_0$ are independent of velocity (whereas its eigenvalues may depend on speed) and that 
$\bm{\beta}-{\bf C}$ is proportional to the identity matrix ${\bf I}$, \cf also \cite[Appendix A]{Ciurylo+Pine2000}.
In principle, a rigorous diagonalization of the line mixing problem is also possible based on velocity dependent eigenvectors
if \eg ${\bf C}={\bf 0}$ and $\bm{\beta}={\bf 0}$, but it will generally require a rather exhaustive 
numerical evaluation of the associated velocity integral, \cf \cite[Sect.~3.2]{Ciurylo+Pine2000}.
Hence, in the following we will assume that the eigenvectors are independent of velocity and that we have a case of Dicke narrowing with uncorrelated hard collisions where
${\bf C}={\bf 0}$ and $\bm{\beta}=\beta{\bf I}$ where $\beta$ is the line-independent frequency of velocity changing 
collisions\footnote{To adhere to standard notation we purposely employ the notation $\beta$ here, but which should cause no confusion with the other definition above where $\beta=1/k_\mrm{B}T$.}.
We recall that the relaxation matrix ${\bf W}$ satisfies the condition of detailed balancing, \ie ${\bf W}\bm{\rho}=\bm{\rho}{\bf W}^\mrm{T}$.
We can then introduce the symmetric matrix  $\bm{\Gamma}=\bm{\rho}^{-1/2}{\bf W}\bm{\rho}^{1/2}$ and start
by diagonalizing the complex symmetric matrix 
\begin{equation}\label{eq:spectdecomp}
\bm{\Gamma}+\iu\bm{\omega}_0={\bf Q}\bm{\Lambda}{\bf Q}^\mrm{T},
\end{equation}
where $\bm{\Lambda}$ is a diagonal matrix of complex eigenvalues $\lambda_n=\gamma_n+\iu\omega_n$ and 
${\bf Q}^{-1}={\bf Q}^\mrm{T}$, \cf \cite[Theorem 4.4.13 on p.~211-212]{Horn+Johnson1985}.
By pre- and post-multiplying \eqref{eq:spectdecomp} with $\bm{\rho}^{1/2}$ and $\bm{\rho}^{-1/2}$, respectively, we can readily see that
\begin{equation}\label{eq:spectdecompW}
{\bf W}+\iu\bm{\omega}_0={\bf A}\bm{\Lambda}{\bf A}^\mrm{-1},
\end{equation}
where ${\bf A}=\bm{\rho}^{1/2}{\bf Q}$ and ${\bf A}^{-1}={\bf Q}^\mrm{T}\bm{\rho}^{-1/2}$.
We can also see that
\begin{multline}
 {\bf W}+\beta{\bf I}-\iu\left(\omega\cdot{\bf I}-\bm{\omega}_0-\bm{k}\cdot\bm{v}\cdot {\bf I}\right) \\
 ={\bf A}\left(\bm{\Lambda}+(\beta-\iu\omega+\iu\bm{k}\cdot\bm{v})\cdot{\bf I}\right){\bf A}^\mrm{-1},
\end{multline}
and hence diagonalize the expression \eqref{eq:Gvomegadef} as
\begin{multline}\label{eq:Gvomegasol}
{\bf G}(\bm{v},\omega)={\bf A}\left(\bm{\Lambda}+(\beta-\iu\omega+\iu\bm{k}\cdot\bm{v})\cdot{\bf I}\right)^{-1}{\bf A}^\mrm{-1}  \\
={\bf A}\cdot\mrm{diag}\left\{ \frac{1}{\gamma_n+\beta-\iu(\omega-\omega_n)+\iu\bm{k}\cdot\bm{v}}\right\}{\bf A}^\mrm{-1}.
\end{multline}

As already mentioned above, we will now assume that ${\bf A}$ is independent of velocity. 
It follows then from \eqref{eq:Gomegadef} and \eqref{eq:Gvomegasol} that 
\begin{equation}\label{eq:Gomegaexpr1}
{\bf G}(\omega)=\int_{\bm{v}}f(\bm{v}){\bf G}(\bm{v},\omega)\mrm{d}\bm{v}
={\bf A}{\bf D}(\omega){\bf A}^\mrm{-1},
\end{equation}
where 
\begin{multline}\label{eq:Domega}
{\bf D}(\omega)=\mrm{diag}\left\{ \int_{\bm{v}} \frac{f(\bm{v})\mrm{d}\bm{v}}{\gamma_n+\beta-\iu(\omega-\omega_n)+\iu\bm{k}\cdot\bm{v}}\right\} \\
=\mrm{diag}\left\{ \frac{\sqrt{\pi}}{k\widetilde{v}}w\left(\frac{\omega-\omega_n+\iu(\gamma_n+\beta)}{k\widetilde{v}}\right) \right\}
\end{multline}
and where the last line is valid when the eigenvalues $\lambda_n$ are independent of velocity.
Here, $w(z)$ denotes the Faddeeva function as defined in Appendix \ref{sect:integrals} and \eqref{eq:ComegaDoppler6} has been used in the last step,
\cf also \cite[p.~381]{Ciurylo+Pine2000}.

Based on \eqref{eq:Ctildeomegasol} with ${\bf C}={\bf 0}$ and $\bm{\beta}=\beta{\bf I}$ as well as the factorization \eqref{eq:Gomegaexpr1}, 
we can now write the final Fourier-Laplace transform as
\begin{multline}\label{eq:finComegasol}
\widetilde{C}(\omega)=\bm{\mu}^\mrm{T}\left({\bf I}- \beta{\bf G}(\omega) \right)^{-1}{\bf G}(\omega)\bm{\rho}\bm{\mu} \\
=\bm{\mu}^\mrm{T}\left({\bf A}{\bf I}{\bf A}^\mrm{-1}- \beta{\bf A}{\bf D}(\omega){\bf A}^\mrm{-1} \right)^{-1} 
{\bf A}{\bf D}(\omega){\bf A}^\mrm{-1}\bm{\rho}\bm{\mu} \\
=\bm{\mu}^\mrm{T}{\bf A}\left({\bf I}-\beta{\bf D}(\omega) \right)^{-1}{\bf D}(\omega){\bf A}^\mrm{-1}\bm{\rho}\bm{\mu} \\
=\bm{\mu}^\mrm{T}\bm{\rho}^{1/2}{\bf Q}\left({\bf I}-\beta{\bf D}(\omega) \right)^{-1}{\bf D}(\omega){\bf Q}^\mrm{T}\bm{\rho}^{-1/2}\bm{\rho}\bm{\mu} \\
={\bf M}^\mrm{T}{\bf Q}\left({\bf I}-\beta{\bf D}(\omega) \right)^{-1}{\bf D}(\omega){\bf Q}^\mrm{T}{\bf M},
\end{multline}
where ${\bf A}=\bm{\rho}^{1/2}{\bf Q}$, ${\bf A}^{-1}={\bf Q}^\mrm{T}\bm{\rho}^{-1/2}$ and ${\bf M}=\bm{\rho}^{1/2}\bm{\mu}$.
Let us now just briefly return to the general case \eqref{eq:Ctildeomegasol} and realize that the factorization ${\bf G}(\omega)={\bf A}{\bf D}(\omega){\bf A}^\mrm{-1}$
does not help diagonalize the final expression \eqref{eq:finComegasol} unless the matrices ${\bf A}^\mrm{-1}$ and $\bm{\beta}- {\bf C}$ commute, \cf \cite[Appendix A]{Ciurylo+Pine2000}.
We can also see from the analysis above why it is so useful that ${\bf A}$ is independent of velocity, 
facilitating a swift evaluation of the velocity integrals in terms of the Faddeeva function, as illustrated in \eqref{eq:Domega}.
Thus, coming back to our special case with uncorrelated Dicke narrowing where $\bm{\beta}- {\bf C}=\beta{\bf I}$ as in \eqref{eq:finComegasol} above, we can now see that
$\widetilde{C}(\omega)$ can be written as the following sum over individual lines
\begin{equation}\label{eq:Comegalines}
\widetilde{C}(\omega)=\sum_n \frac{a_n^2D_{n}(\omega)}{1-\beta D_n(\omega)},
\end{equation}
where $D_{n}(\omega)$ are the diagonal elements defined by \eqref{eq:Domega}, and
\begin{equation}\label{eq:andef}
a_n={\bf M}^\mrm{T}{\bf q}_n
\end{equation}
where ${\bf q}_n$ is the $n$th column of ${\bf Q}$, \cf also \cite[Eq.~(3.21)]{Ciurylo+Pine2000} and \cite[Eq.~(15)]{Pine1997}.
It is emphasized here that $a_n^2$ is complex and \eqref{eq:Comegalines} takes full line mixing into account.

The effect of Doppler broadening can be ignored by putting the wave number $k=0$ while keeping $\omega$ fixed, which immediately yields
\begin{equation}\label{eq:Domegak0}
D_{n}(\omega)=\frac{1}{\gamma_n+\beta-\iu(\omega-\omega_n)},
\end{equation}
provided that the eigenvalues $\lambda_n=\gamma_n+\iu\omega_n$ are independent of velocity.
However, in this case we can readily see that the solution becomes independent of $\beta$, and hence
\begin{equation}\label{eq:Comegalinesk0}
\widetilde{C}(\omega)=\sum_n \frac{a_n^2}{\gamma_n-\iu(\omega-\omega_n)}.
\end{equation}
It should be emphasized here that in case of speed-dependent collisional width and shift and $k=0$ 
the line shape does in fact depend on parameter $\beta$, \cf \eg \cite{Robert+etal1993}.

\subsection{Rosenkranz parameters}

We consider now the spectral decomposition \eqref{eq:spectdecomp}, or equivalently
\eqref{eq:spectdecompW} where $\bm{\Gamma}=\bm{\rho}^{-1/2}{\bf W}\bm{\rho}^{1/2}$.
We recall that \\
$\left(\bm{\Gamma}+\iu\bm{\omega}_0\right){\bf q}_n=\lambda_n{\bf q}_n$
and $\lambda_n=\gamma_n+\iu\omega_n$.
Assuming that $\bm{\Gamma}=p\widehat{\bm{\Gamma}}$ where $p$ is pressure, 
the Rosenkranz parameters are given by the following first order approximations
\begin{equation}\label{eq:Rosenkranzlambdan}
\lambda_n=p\widehat{\Gamma}_{nn}+\iu\omega_{0n},
\end{equation}
and
\begin{equation}\label{Rosenkranzqn}
q_{kn}=\left\{\begin{array}{ll}
1 & k=n \vspace{0.2cm} \\
\displaystyle\frac{\iu p\widehat{\Gamma}_{kn}}{\omega_{0k}-\omega_{0n}} & k\neq n,
\end{array}\right.
\end{equation}
where $q_{kn}$ are the elements of ${\bf q}_n$, \cf \cite[(A.4) and (A.5)]{Rosenkranz1975}.
It is noted that this analysis can be performed already in velocity space if the eigenvalues $\lambda_n$ depend on $\bm{v}$.
Based on \eqref{eq:andef} and \eqref{Rosenkranzqn}, we can now find that
\begin{equation}\label{eq:anfirstorder}
a_n=M_n+\iu p\sum_{k\neq n}M_k\frac{\widehat{\Gamma}_{kn}}{\omega_{0k}-\omega_{0n}}
\end{equation}
and to first order in $p$ we can also derive the following expression
\begin{multline}\label{eq:squarean}
a_n^2=M_n^2+2\iu p M_n\sum_{k\neq n}\frac{M_k\widehat{\Gamma}_{kn}}{\omega_{0k}-\omega_{0n}} \\
=\rho_n\mu_n^2+2\iu p \rho_n\mu_n\sum_{k\neq n}\frac{\mu_k\widehat{W}_{kn}}{\omega_{0k}-\omega_{0n}},
\end{multline}
where we have also employed the definitions $M_n=\rho_n^{1/2}\mu_n$ and $\widehat{\Gamma}_{kn}=\rho_k^{-1/2}\widehat{W}_{kn}\rho_n^{1/2}$, \cf \cite[Eq.~(2) and (3)]{Rosenkranz1975}.

\section{A modified projection approach to line mixing}\label{sect:ModProj}
\subsection{The basic projection based method}
The basic projection based strong collsion (SC) method has been proposed 
in \eg \cite{Dokuchaev+etal1982,Bulanin+etal1984,Filippov+Tonkov1993,Tonkov+etal1996,Filippov+etal2002,Tonkov+Filippov2003}
and is briefly summarized below.
Within this strong collision model it is assumed that the relaxation time $\tau_s$ is equal to the mean duration between successive collisions for 
any rotational state of the absorbing molecule \cite{Filippov+Tonkov1993}.
It is furthermore assumed that $\omega_{0n}\tau_\mrm{s} \ll 1$, and the corresponding collision frequency is defined as $v_\mrm{s}=\tau_\mrm{s}^{-1}$.
The symmetrized relaxation matrix $\bm{\Gamma}$ can now be defined in terms of a projector ${\bf I}-\tau_\mrm{s}\bm{\Gamma}$
that restores thermal equilibrium at the characteristic time $\tau_\mrm{s}$, and hence 
\begin{equation}\label{eq:GammaSC}
\bm{\Gamma}=v_\mrm{s}\left({\bf I}-\frac{{\bf M}{\bf M}^\mrm{T}}{{\bf M}^\mrm{T}{\bf M}}\right),
\end{equation}
where ${\bf M}=\bm{\rho}^{1/2}\bm{\mu}$, \cf \eg \cite[Eq.~(A18)]{Bulanin+etal1984}, \cite[Eq.~(13)]{Tonkov+Filippov2003} and \cite[p.~113]{Filippov+Tonkov1993}.
The corresponding unsymmetric relaxation matrix is then given by 
\begin{equation}
{\bf W}=\bm{\rho}^{1/2}\bm{\Gamma}\bm{\rho}^{-1/2}=v_\mrm{s}\left({\bf I}-\frac{\bm{\rho}\bm{\mu}\bm{\mu}^\mrm{T}}{\bm{\mu}^\mrm{T}\bm{\rho}\bm{\mu}}\right),
\end{equation}
and which thus satisfies the condition of detailed balancing ${\bf W}\bm{\rho}=\bm{\rho}{\bf W}^\mrm{T}$.
It is furthermore recalled here that $C(0)=\bm{\mu}^\mrm{T}\bm{\rho}\bm{\mu}={\bf M}^\mrm{T}{\bf M}=\sum_n S_n$
where the line strengths are given by $S_n=M_n^2$ and $M_n=\rho_n^{1/2}\mu_n$.

The line width parameter $v_\mrm{s}$ can now be chosen so that  the theoretical or experimental line widths $\gamma_{0n}$
are suitably approximated by the diagonal elements $W_{nn}=\Gamma_{nn}=v_\mrm{s}(1-S_n/C(0))$.
To this end, the following formula 
\begin{equation}\label{eq:vssol2}
v_\mrm{s}=\frac{\displaystyle\sum_n \gamma_{0n} S_n}{\displaystyle\sum_n S_n},
\end{equation}
has been suggested in \eg \cite[p.~113]{Filippov+Tonkov1993} and \cite[Eq.~(16)]{Tonkov+Filippov2003}.

The strong collision (SC) model above has been validated against experimental data and compared to 
an improved technique referred to as adjustable branch coupling (ABC) in \eg \cite{Tonkov+etal1996,Tonkov+Filippov2003,Domanskaya+etal2004}.
It has been demonstrated that both the SC and ABC methods are able to provide significantly better predictions of the far wing 
behavior of an absorbing gas in comparison to a simple sum of isolated Lorentzian lines.
The ABC method is however more sophisticated as it requires the subdivision of lines into isolated branches, but it is also able to provide better predictions
 as it employs one additional free parameter (interbranch interaction) to match the model to experimental data.
 In our approach here we wish to retain the simplicity of the SC projection method to line coupling described above,
 while at the same time reinstalling the high-resolution aspects of some standard isolated line models.
 This will be the topic of the sections that follow.
 
\subsection{Perturbation}
We will now aim to improve the simple strong collision (SC) projection method above by introducing the slightly perturbed model
\begin{equation}\label{eq:GammaSCmod}
{\Gamma}_{kn}=\left\{\begin{array}{ll}
\displaystyle \gamma_{0n}(v)+\iu\delta_{0n}(v) & k=n, \vspace{0.2cm} \\
\displaystyle -v_\mrm{s} \frac{M_kM_n}{C(0)} & k\neq n,
\end{array}\right.
\end{equation}
where the diagonal elements in \eqref{eq:GammaSC} have been replaced by any (theoretical or experimental)
presumably more accurate and possibly even speed dependent broadening and shift parameters $\gamma_{0n}(v)+\iu\delta_{0n}(v)$.
The same off-diagonal elements as in \eqref{eq:GammaSC} are retained
as a model of the line coupling transfer rates. 
It is noted that the unsymmetric relaxation matrix ${\bf W}=\bm{\rho}^{1/2}\bm{\Gamma}\bm{\rho}^{-1/2}$ is still satisfying the condition of detailed balancing, just as before.
The parameter $v_\mrm{s}$ should be chosen to maintain as much as possible
of the thermalizing projector property, but could also be treated as
an empirical parameter that can readily be adjusted to match experimental data.

The modification introduced in \eqref{eq:GammaSCmod} means that we have now added to \eqref{eq:GammaSC} the diagonal perturbation matrix 
\begin{equation}\label{eq:Pdef}
{\bf P}=\mrm{diag}\{\gamma_{0n}+\iu\delta_{0n}-v_\mrm{s}(1-S_n/C(0))\},
\end{equation}
and consequently the perturbed matrix \eqref{eq:GammaSCmod} does no longer correspond to a rigorous projector. 
However, the perturbed model does provide a useful compromise 
in the sense that the rigorous projector is only slightly relaxed in favor of providing two new features: 
An accurate modeling close to the line centers as well as an accurate modeling in the far wings, the latter being achieved by fine tuning the parameter $v_\mrm{s}$.

Let us now briefly discuss the properties of the two approximate projectors ${\bf I}-\tau_\mrm{s}\bm{\Gamma}$ and $\tau_\mrm{s}\bm{\Gamma}$,
the former ideally being a projector onto the one-dimensional space parallel to ${\bf M}$ and the latter orthogonal to ${\bf M}$. Let us now consider a vector
${\bf x}={\bf x}_\parallel+{\bf x}_\perp$ being correspondingly represented in line space, and where 
\begin{equation}\label{eq:Linespacedecomp}
\left({\bf I}-\tau_\mrm{s}\bm{\Gamma}\right){\bf x}={\bf x}_\parallel+\left({\bf I}-\tau_\mrm{s}\bm{\Gamma}\right){\bf x}_\perp-\tau_\mrm{s}\bm{\Gamma}{\bf x}_\parallel.
\end{equation}
The first term on the right-hand side of \eqref{eq:Linespacedecomp} is what the projector is aiming for. 
It is now assumed that the state of the system involving molecular collisions 
is always relatively close to thermal equilibrium, and hence that the modulus of the vector ${\bf x}_\parallel$ is much larger than the modulus of ${\bf x}_\perp$. 
The perturbed matrix  ${\bf I}-\tau_\mrm{s}\bm{\Gamma}$ is furthermore an approximate projector almost ortogonal to ${\bf x}_\perp$, all of which now makes the
second term negligible. Hence, it is the vanishing of the last term in \eqref{eq:Linespacedecomp} that is the most important for maintaining the required projection property.
To this end, it is noted that the required orthogonality property $\bm{\Gamma}{\bf M}={\bf 0}$ can also be interpreted as a sum rule 
valid for a rigid rotor where the line coupling depend only on its rotational states, \cf \cite[Eq.~(A9)]{Bulanin+etal1984}, 
\cite[p.~113]{Filippov+Tonkov1993} and \cite[Eq.~(4)]{Tonkov+Filippov2003}.
In the present context this means that we should now choose the parameter $v_\mrm{s}$ to
minimize the least squares norm of the vector ${\bf P}{\bf M}$, yielding
\begin{equation}\label{eq:vssol1}
v_\mrm{s}^\mrm{ls}=\frac{\displaystyle\sum_n \gamma_{0n} S_n \left(1-\frac{S_n}{C(0)} \right)}{\displaystyle\sum_n S_n\left(1-\frac{S_n}{C(0)} \right)^2 },
\end{equation}
where $C(0)=\sum_n S_n$. It may be noticed that the expression \eqref{eq:vssol1} also corresponds to an $S_n$-weighted least squares solution
to minimize the error $v_\mrm{s}(1-S_n/C(0))-\gamma_{0n}$ related to \eqref{eq:GammaSC}, and that \eqref{eq:vssol2} is obtained if the ratio $S_n/C(0)$ is neglected.
Notably, if $\gamma_{0n}$ is a speed dependent parameter, we would use in \eqref{eq:vssol1} 
either an average $\bar{\gamma}_{0n}=\int_{\bm{v}}f(\bm{v})\gamma_{0n}\mrm{d}\bm{v}$, or we would evaluate $\gamma_{0n}$ at the
most probable speed $\widetilde{v}$ to obtain a speed independent parameter $v_\mrm{s}$.

In practice, we have found that it may be useful to make a very small fine tuning of \eqref{eq:vssol1} to slightly increase the line coupling transfer rates 
for a better match to measurement data. Hence, we may choose $v_\mrm{s}=cv_\mrm{s}^\mrm{ls}$ where $c$ is a constant very close to 1 ($c=1.005$ in our numerical 
example below for $\mrm{CO}_2$ in the $\nu_3$-band).
This fine tuning of $v_\mrm{s}$ has insignificant effect on the absorption close to the line centers, but it can provide an appropriate correction of the far wing behavior.
%Obviously, the parameter $c$ can readily be adjusted to match the experimental data of any specific species and ro-vibrational bands of interest.

\subsection{Rosenkranz parameters}
Since the matrix ${\bf W}$ is assumed to be linear with pressure $p$ and $\bm{\Gamma}=p\widehat{\bm{\Gamma}}$, 
we introduce now also the notation $\Gamma_{nn}=\gamma_{0n}+\iu\delta_{0n}=p(\widehat{\gamma}_{0n}+\iu\widehat{\delta}_{0n})$ 
and $v_\mrm{s}=p\widehat{v}_\mrm{s}$.
By following \eqref{eq:Rosenkranzlambdan} through \eqref{eq:squarean}
and to the first order in $p$, the corresponding Rosenkranz parameters are now obtained as 
\begin{equation}\label{eq:RosenkranzparamsSC}
\left\{\begin{array}{l}
\gamma_n =p\widehat{\gamma}_{0n}, \vspace{0.2cm} \\ 
\omega_n=\omega_{0n}+p\widehat{\delta}_{0n}, 
\end{array}\right.
\end{equation}
as well as
\begin{equation}\label{eq:anfirstorderSC}
a_n=M_n+\frac{\iu p\widehat{v}_\mrm{s}M_n}{C(0)}\sum_{k\neq n}\frac{S_k }{\omega_{0n}-\omega_{0k}}
\end{equation}
and
\begin{equation}\label{eq:ansquarefirstorderSC}
a_n^2=M_n^2+\frac{2\iu p\widehat{v}_\mrm{s}M_n^2}{C(0)}\sum_{k\neq n}\frac{S_k }{\omega_{0n}-\omega_{0k}}.
\end{equation}
The spectral function $\widetilde{C}(\omega)$ can then finally be calculated as in \eqref{eq:Comegalines} where $D_n(\omega)$ has been defined in \eqref{eq:Domega}.

In the case with velocity independent parameters the only prior knowledge required for the computation of \eqref{eq:RosenkranzparamsSC},
\eqref{eq:anfirstorderSC} and \eqref{eq:ansquarefirstorderSC}
are the center frequencies $\omega_{0n}$, the line widths $\gamma_{0n}$ and shifts $\delta_{0n}$ and the line strengths $S_n$. These are parameters that are readily available for 
a large variety of species in spectroscopic databases such as \eg HITRAN
\cite{HITRANorg,HITRANdefs,Rothman+etal1998,Simeckova+etal2006,Gordon+etal2022}.

The case with speed dependent diagonal elements $\Gamma_{nn}(v)$ can also be treated under the Rosenkranz approximation.
We may consider the following quadratic model for the speed dependent pressure broadening and shifts
\begin{equation}\label{eq:quadraticGammaDelta1}
\gamma_{0n}(v)+\iu\delta_{0n}(v)=C_{0n}+C_{2n}\left(v^2/\widetilde{v}^2-3/2\right)
\end{equation}
where $C_{0n}$ and $C_{2n}$ are complex valued parameters which are proportional to pressure, \cf \eg the Hartmann-Tran profile \cite{Ngo+etal2013,Tennyson+etal2014}. 
It is furthermore noticed here that $\langle v^2/\widetilde{v}^2 \rangle=\int_{\bm{v}}f(\bm{v})\left(v^2/\widetilde{v}^2\right)\mrm{d}\bm{v}=3/2$.
The defining integral in \eqref{eq:Domega} then becomes
\begin{multline}\label{eq:DomegaHTPex}
D_n(\omega)= \int_{\bm{v}} \frac{f(\bm{v})\mrm{d}\bm{v}}{\gamma_n+\beta-\iu(\omega-\omega_n)+\iu\bm{k}\cdot\bm{v}}\\
= \int_{\bm{v}} \frac{f(\bm{v})\mrm{d}\bm{v}}{C_{0n}+C_{2n}\left(v^2/\widetilde{v}^2-3/2\right)+\beta-\iu(\omega-\omega_{0n})+\iu\bm{k}\cdot\bm{v}},
\end{multline}
and which can be expressed explicitly in terms of two Faddeeva function evaluations as explained in \cite[Appendix A and B]{Ngo+etal2013},
\cf also \cite[Eq.~(3.30)]{Ciurylo+Pine2000}.
The parameter $a_n$ can now be computed in the same way as before in \eqref{eq:anfirstorderSC} or \eqref{eq:ansquarefirstorderSC}
and then finally $\widetilde{C}(\omega)$ as in \eqref{eq:Comegalines}.

\subsection{Exact solution}
Following the ideas presented in \cite[Eq.~(10)-(12)]{Tonkov+etal1996} and \cite[Eq.~(14)-(15)]{Tonkov+Filippov2003},
it is possible to develop a simple closed form solution to \eqref{eq:Ctildeomegasol}
for the case with uncorrelated collisions without velocity changes and where $\bm{\Gamma}$
is given by the modified projection model \eqref{eq:GammaSCmod}.
Hence, in this case we have ${\bf C}={\bf 0}$ and $\bm{\beta}=\bm{0}$, and \eqref{eq:Gvomegadef} and \eqref{eq:Ftildeeq2} yield
\begin{multline}\label{eq:Comegaexact}
\widetilde{C}(\bm{v},\omega)=\bm{\mu}^\mrm{T}\widetilde{\bf F}(\bm{v},\omega) \\
=\bm{\mu}^\mrm{T}\left( {\bf W}-\iu\left(\omega\cdot{\bf I}-\bm{\omega}_0-\bm{k}\cdot\bm{v}\cdot {\bf I}\right)\right)^{-1} 
f(\bm{v})\bm{\rho}\bm{\mu}.
\end{multline}
By employing ${\bf W}=\bm{\rho}^{1/2}\bm{\Gamma}\bm{\rho}^{-1/2}$ and ${\bf M}=\bm{\rho}^{1/2}\bm{\mu}$ we obtain the symmetrized form
\begin{equation}\label{eq:Comegaexact2}
\widetilde{C}(\bm{v},\omega)
={\bf M}^\mrm{T}\left( \bm{\Gamma}-\iu\left(\omega\cdot{\bf I}-\bm{\omega}_0-\bm{k}\cdot\bm{v}\cdot {\bf I}\right)\right)^{-1} {\bf M}f(\bm{v}),
\end{equation}
and by inserting \eqref{eq:GammaSCmod} we obtain
\begin{multline}
\widetilde{C}(\bm{v},\omega) 
={\bf M}^\mrm{T}\left({\bf P}+v_\mrm{s}\left({\bf I}-\frac{{\bf M}{\bf M}^\mrm{T}}{C(0)}\right) \right. \\
\left. -\iu\left(\omega\cdot{\bf I}-\bm{\omega}_0-\bm{k}\cdot\bm{v}\cdot {\bf I}\right) \right)^{-1} {\bf M}f(\bm{v}),
\end{multline}
where ${\bf P}=\mrm{diag}\{\gamma_{0n}+\iu\delta_{0n}-v_\mrm{s}(1-S_n/C(0))\}$.
Now, we introduce the matrices
\begin{equation}
{\bf D}^{-1}=\mrm{diag}\left\{\gamma_{0n}+\iu\delta_{0n}+v_\mrm{s}\frac{S_n}{C(0)}-\iu\left(\omega-\omega_{0n}- \bm{k}\cdot\bm{v}\right) \right\},
\end{equation}
and 
\begin{equation}
{\bf E}^{-1}=-\frac{v_\mrm{s}}{C(0)}
\end{equation}
and write
\begin{equation}
\widetilde{C}(\bm{v},\omega)={\bf M}^\mrm{T}\left( {\bf D}^{-1} + {\bf M}{\bf E}^{-1}{\bf M}^\mrm{T} \right)^{-1} {\bf M}f(\bm{v}).
\end{equation}
By making use of the matrix inversion lemma \cite[p.~30]{Mendel1987}, it can readily be seen that
the exact solution in velocity space is given by
\begin{multline}
\widetilde{C}(\bm{v},\omega) \\ ={\bf M}^\mrm{T}\left({\bf D}-{\bf D}{\bf M}\left({\bf M}^\mrm{T}{\bf D}{\bf M}+{\bf E} \right)^{-1}{\bf M}^\mrm{T}{\bf D}  \right){\bf M}f(\bm{v}).
\end{multline}
After some algebra, it is found that
\begin{equation}\label{eq:extactCvomega} 
\widetilde{C}(\bm{v},\omega)=\frac{{\bf M}^\mrm{T}{\bf D}{\bf M}f(\bm{v})}{1+{\bf E}^{-1}{\bf M}^\mrm{T}{\bf D}{\bf M}}=
\frac{\widetilde{C}_1(\bm{v},\omega)f(\bm{v})}{1-\frac{v_\mrm{s}}{C(0)}\widetilde{C}_1(\bm{v},\omega)}
\end{equation}
where
\begin{multline}\label{eq:extactC1vomega}
\widetilde{C}_1(\bm{v},\omega)={\bf M}^\mrm{T}{\bf D}{\bf M} \\
=\sum_n \frac{S_n}{\gamma_{0n}+v_\mrm{s}\frac{S_n}{C(0)}-\iu\left(\omega-\omega_{0n}-\delta_{0n} - \bm{k}\cdot\bm{v}\right)},
\end{multline}
and where $S_n=M_n^2$ and $C(0)=\sum_n S_n$.
The total Fourier-Laplace transform $\widetilde{C}(\omega)$ is finally obtained by integrating over velocity space as
\begin{equation}\label{eq:exactComega}
\widetilde{C}(\omega)=\int_{\bm{v}}\widetilde{C}(\bm{v},\omega)\mrm{d}\bm{v}.
\end{equation}

In the case when the Doppler effect can be neglected we can set $k=0$ in \eqref{eq:extactC1vomega},
and if the broadening and shift parameters $\gamma_{0n}+\iu\delta_{0n}$ are furthermore independent of velocity then
\eqref{eq:extactCvomega} gives after integration
\begin{equation}\label{eq:extactComegakeq0}
\widetilde{C}(\omega)=\frac{\widetilde{C}_1(\omega)}{1-\frac{v_\mrm{s}}{C(0)}\widetilde{C}_1(\omega)}
\end{equation}
where
\begin{equation}\label{eq:extactC1omegakeq0}
\widetilde{C}_1(\omega)
=\sum_n \frac{S_n}{\gamma_{0n}+v_\mrm{s}\frac{S_n}{C(0)}-\iu\left(\omega-\omega_{0n}-\delta_{0n}\right)}.
\end{equation}

Unfortunately, the function \eqref{eq:extactCvomega}  can not readily be integrated over velocity space 
based on \eqref{eq:extactC1vomega}  when Doppler broadening is present and $k\neq 0$. 
However, if we can assume that the Boltzmann factor $\eu^{-(v/\widetilde{v})^2}$ is very narrow in comparison to the variations in $\widetilde{C}_1(\bm{v},\omega)$,
 \ie if $k\widetilde{v}/\gamma_{0n}\ll 1$, 
then we may approximate \eqref{eq:exactComega} by applying the $f(\bm{v})$-weighted integration (averaging) separately to the numerator and the denominator 
of \eqref{eq:extactCvomega}, respectively.
Assuming once again that $\gamma_{0n}+\iu\delta_{0n}$ are independent of velocity this will then yield a result of the same form as in \eqref{eq:extactComegakeq0} where
\begin{multline}\label{eq:approxC1omegasol}
\widetilde{C}_1(\omega)=\int_{\bm{v}}\widetilde{C}_1(\bm{v},\omega)f(\bm{v})\mrm{d}\bm{v} \\
=\frac{\sqrt{\pi}}{k\widetilde{v}}\sum_n S_n w\left(\frac{\omega-\omega_{0n}-\delta_{0n}
+\iu \left(\gamma_{0n}+v_\mrm{s}\frac{S_n}{C(0)}\right)}{k\widetilde{v}} \right)
\end{multline}
and where $w(z)$ is the Faddeeva function based on the integral identity \eqref{eq:ComegaDoppler6}.
Eventhough \eqref{eq:approxC1omegasol} does not provide a rigorous calculation of \eqref{eq:exactComega},
it is similar to the results obtained in \cite[Eq.~(9)]{Filippov+etal2002}, 
and it has the correct asymptotics as $k\rightarrow 0$ in accordance with \eqref{eq:extactC1omegakeq0}.
However, perhaps a more rigorous alternative for including the Doppler effect is then to employ the diagonalization \eqref{eq:Domega}
and \eqref{eq:Comegalines} together with the Rosenkranz parameters 
\eqref{eq:RosenkranzparamsSC} and \eqref{eq:anfirstorderSC}, and which is also providing
the option to include Dicke narrowing with parameter $\beta$.

\section{Wavenumber domain and the fluctuation-dissipation theorem}\label{sect:wavenumber}

Let us now summarize the results of the previous section by writing the expressions in the wavenumber domain 
while at the same time maintaining the possibility to include the influence of the fluctuation-dissipation theorem as in \eqref{eq:sigmaaexpr}. 
The latter will finally be achieved by incorporating an appropriate scaling of the line strengths which are retrieved from the HITRAN database.
To do this carefully we will consider here the expressions in SI-units for simplicity and then finally execute the computations in reciprocal centimeters as usual.

\subsection{Wavenumber domain}
The wavenumber domain is introduced via the substitution 
$\omega=2\pi\mrm{c}_0\nu$ where 
$\nu=\lambda^{-1}$ is the wavenumber and $\lambda$ the wavelength of the radiation. 
We have thus $I(\nu)=2\pi\mrm{c}_0 I(\omega)$ and
\begin{equation}
\int I(\omega)\mrm{d}\omega=\int I(\nu)\mrm{d}\nu=C(0)=\sum_n S_n,
\end{equation}
where the SI-units of the correlation function $C(t)$ as well as the line strengths $S_n$ are given in \unit{A^2s^2m^2}.
The corresponding Fourier-Laplace transform is similarly defined so that $\widetilde{C}(\nu)=2\pi\mrm{c}_0\widetilde{C}(\omega)$
and $I(\nu)=\frac{1}{\pi}\Re\{\widetilde{C}(\nu)\}$. 

We introduce now the following parameter scaling
\begin{equation}
\left[ \gamma_{0n}, \delta_{0n}, \gamma_n, v_\mrm{s}, \beta, \gamma_\mrm{D} \right]
=2\pi\mrm{c}_0\left[ \gamma_{0n}^\prime, \delta_{0n}^\prime, \gamma_n^\prime, v_\mrm{s}^\prime, \beta^\prime, \gamma_\mrm{D}^\prime \right],
\end{equation}
where the unprimed parameters refer to the frequency domain and the primed parameters to the wavenumber domain.
Here, we are furthermore introducing the Doppler half-width parameter $\gamma_\mrm{D}=k\widetilde{v}\sqrt{\ln 2}$ so that
\begin{equation}
\gamma_\mrm{D}^\prime
=\frac{\nu\widetilde{v}\sqrt{\ln 2}}{\mrm{c}_0},
\end{equation}
where $k=2\pi\nu$. For notational convenience it is also natural to write $\omega_{0n}=2\pi\mrm{c}_0\nu_{0n}$ and $\omega_{n}=2\pi\mrm{c}_0\nu_{n}$.
The Fourier-Laplace transform in \eqref{eq:Comegalines} now becomes
\begin{equation}\label{eq:Cnulines}
\widetilde{C}(\nu)
=\sum_n \frac{a_n^2D_{n}(\nu)}{1-\beta^\prime D_n(\nu)},
\end{equation}
where 
\begin{multline}\label{eq:Dnnu}
D_n(\nu)=2\pi\mrm{c}_0 D_n(\omega) \\
=\int_{\bm{v}} \frac{f(\bm{v})\mrm{d}\bm{v}}{\gamma_n^\prime+\beta^\prime-\iu(\nu-\nu_n)+\iu\frac{\gamma_\mrm{D}^\prime}{\widetilde{v}\sqrt{\ln 2}}\widehat{\bm{k}}\cdot\bm{v}} \\
=\frac{\sqrt{\pi}}{\gamma_\mrm{D}^\prime/\sqrt{\ln 2}}w\left(\frac{\nu-\nu_n+\iu(\gamma_n^\prime+\beta^\prime)}{\gamma_\mrm{D}^\prime/\sqrt{\ln 2}}\right),
\end{multline}
and where the last line is valid when the eigenvalues $\lambda_n^\prime=\gamma_n^\prime+\iu\nu_n$ are independent of velocity, \cf \eqref{eq:Domega}.
Without Doppler broadening for $k=0$, we employ instead \eqref{eq:Domegak0} to yield
\begin{equation}\label{eq:Dnuk0}
D_{n}(\nu)=\frac{1}{\gamma_n^\prime+\beta^\prime-\iu(\nu-\nu_n)},
\end{equation}
where $\beta^\prime$ is again redundant as in \eqref{eq:Comegalinesk0}, yielding
\begin{equation}\label{eq:Cnulinesk0}
\widetilde{C}(\nu)=\sum_n \frac{a_n^2}{\gamma_n^\prime-\iu(\nu-\nu_n)}.
\end{equation}

We can see from the definition made in \eqref{eq:andef} 
that the parameter $a_n={\bf M}^\mrm{T}{\bf q}_n$ used in \eqref{eq:Cnulines} and \eqref{eq:Cnulinesk0}
is invariant under the substitution $\omega=2\pi\mrm{c}_0\nu$.
In particular, the spectral decomposition $\bm{\Gamma}+\iu\bm{\omega}_0={\bf Q}\bm{\Lambda}{\bf Q}^\mrm{T}$ 
which was defined in \eqref{eq:spectdecomp} now becomes
$\bm{\Gamma}^\prime+\iu\bm{\nu}_0={\bf Q}\bm{\Lambda}^\prime{\bf Q}^\mrm{T}$ where
$\bm{\Gamma}^\prime=\bm{\Gamma}/2\pi\mrm{c}_0$, $\bm{\nu}_0=\mrm{diag}\{\nu_{0n}\}$ and 
$\lambda_n^\prime=\lambda_n/2\pi\mrm{c}_0=\gamma_n^\prime+\iu\nu_n$.
Thus, the eigenvalues $\lambda_n$ are scaled as the frequency parameters, but the eigenvectors $\bm{q}_n$ are dimensionfree invariants.
If there is no line coupling we have ${\bf Q}={\bf I}$, $a_n=M_n$ and $a_n^2=S_n$.

The modified projection method based on the perturbed relaxation matrix \eqref{eq:GammaSCmod} is now given by
\begin{equation}\label{eq:GammaSCmodnu}
{\Gamma}_{kn}^\prime=\left\{\begin{array}{ll}
\displaystyle \gamma_{0n}^\prime(v)+\iu\delta_{0n}^\prime(v) & k=n, \vspace{0.2cm} \\
\displaystyle -v_\mrm{s}^\prime \frac{M_kM_n}{C(0)} & k\neq n,
\end{array}\right.
\end{equation}
and where $\gamma_{0n}^\prime+\iu\delta_{0n}^\prime=p(\widehat{\gamma}_{0n}^\prime+\iu\widehat{\delta}_{0n}^\prime)$ 
and $v_\mrm{s}^\prime=p\widehat{v}_\mrm{s}^\prime$ where $p$ is pressure.
Assuming that the broadening and shift parameters $\gamma_{0n}^\prime+\iu\delta_{0n}^\prime$ are furthermore independent of velocity,
the design parameter $\widehat{v}_\mrm{s}^\prime$ can initially be chosen as in \eqref{eq:vssol1}, which now becomes
\begin{equation}\label{eq:vssol1nu}
\widehat{v}_\mrm{s}^{\mrm{ls}\prime}=\frac{\displaystyle\sum_n \widehat{\gamma}_{0n}^\prime S_n \left(1-\frac{S_n}{C(0)} \right)}{\displaystyle\sum_n S_n\left(1-\frac{S_n}{C(0)} \right)^2 },
\end{equation}
where $C(0)=\sum_n S_n$ and $M_n=\sqrt{S_n}$.
%In order to fine tune the far wing accuracy of this model the parameter $\widehat{v}_\mrm{s}^\prime$ can now be chosen as $\widehat{v}_\mrm{s}^\prime=c\widehat{v}_\mrm{s}^{\mrm{ls}\prime}$
%where $c$ is a constant very close to one.

\subsection{Rosenkranz parameters}
The general first order Rosenkranz parameters are now obtained from \eqref{eq:Rosenkranzlambdan} through \eqref{eq:anfirstorder}  as
\begin{equation}\label{eq:Rosenkranzlambdannu}
\lambda_n^\prime=p\widehat{\Gamma}_{nn}^\prime+\iu\nu_{0n},
\end{equation}
as well as
\begin{equation}\label{eq:anfirstordernu}
a_n=M_n+\iu p\sum_{k\neq n}M_k\frac{\widehat{\Gamma}_{kn}^\prime}{\nu_{0k}-\nu_{0n}}
\end{equation}
where $\widehat{\Gamma}_{kn}^\prime=\widehat{\Gamma}_{kn}/2\pi\mrm{c}_0$.
The corresponding Rozenkranz parameters based on the modified projection method \eqref{eq:GammaSCmod} are 
given by \eqref{eq:RosenkranzparamsSC} and \eqref{eq:anfirstorderSC} 
and which transforms to the wavenumber domain as
\begin{equation}\label{eq:RosenkranzparamsSCnu}
\left\{\begin{array}{l}
\gamma_n^\prime =p\widehat{\gamma}_{0n}^\prime, \vspace{0.2cm} \\ 
\nu_n=\nu_{0n}+p\widehat{\delta}_{0n}^\prime, 
\end{array}\right.
\end{equation}
and
\begin{equation}\label{eq:anfirstorderSCnu}
a_n=M_n+\frac{\iu p\widehat{v}_\mrm{s}^\prime M_n}{C(0)}\sum_{k\neq n}\frac{S_k }{\nu_{0n}-\nu_{0k}}.
\end{equation}

\subsection{Exact solution}
In the case when there is no Doppler broadening ($k=0$)
and if the broadening and shift parameters $\gamma_{0n}^\prime+\iu\delta_{0n}^\prime$ are furthermore independent of velocity then
the exact solution in \eqref{eq:extactComegakeq0} and \eqref{eq:extactC1omegakeq0} becomes
\begin{equation}\label{eq:extactCnukeq0}
\widetilde{C}(\nu)=\frac{\widetilde{C}_1(\nu)}{1-\frac{v_\mrm{s}^\prime}{C(0)}\widetilde{C}_1(\nu)}
\end{equation}
where
\begin{equation}\label{eq:extactC1nukeq0}
\widetilde{C}_1(\nu)
=\sum_n \frac{S_n}{\gamma_{0n}^\prime+v_\mrm{s}^\prime\frac{S_n}{C(0)}-\iu\left(\nu-\nu_{0n}-\delta_{0n}^\prime\right)}.
\end{equation}
Similar expressions can be derived with regard to \eqref{eq:approxC1omegasol}.

\subsection{Basic projection method}
As for a comparison, based on the original unperturbed strong collision model \eqref{eq:GammaSC} and  \eqref{eq:vssol2} , we will use instead
\begin{equation}\label{eq:vssol2nu}
\widehat{v}_\mrm{s}^\prime=\frac{\displaystyle\sum_n \widehat{\gamma}_{0n}^\prime S_n}{\displaystyle\sum_n S_n },
\end{equation}
and in the case with no Doppler broadening ($k=0$) and an exact solution, we will employ \eqref{eq:extactCnukeq0}
together with
\begin{equation}\label{eq:extactC1nukeq0SC}
\widetilde{C}_1(\nu)
=\sum_n \frac{S_n}{v_\mrm{s}^\prime-\iu\left(\nu-\nu_{0n}\right)},
\end{equation} 
as suggested in \cite[Eq.~(10)-(12)]{Tonkov+etal1996} and \cite[Eq.~(14)-(15)]{Tonkov+Filippov2003}.
Thus, we can now observe here once again the main difference between the basic and the modified projection method being that the
line width parameter $v_\mrm{s}^\prime$ in \eqref{eq:extactC1nukeq0SC} is a constant whereas $\gamma_{0n}^\prime$ in \eqref{eq:extactC1nukeq0} depends on the line index $n$.
As for the Rosenkranz approximation based on the unperturbed model \eqref{eq:GammaSC}, the only difference is with the computation of eigenvalues where 
\begin{equation}\label{eq:RosenkranzparamsSCnuSC}
\left\{\begin{array}{l}
\gamma_n^\prime =p\widehat{v}_\mrm{s}^\prime(1-S_n/C(0)), \vspace{0.2cm} \\ 
\nu_n=\nu_{0n},
\end{array}\right.
\end{equation}
instead of \eqref{eq:RosenkranzparamsSCnu}.
The other relations \eqref{eq:Cnulines} through \eqref{eq:Cnulinesk0}
and \eqref{eq:anfirstorderSCnu} are obtained as above.

\subsection{The absorption coefficient}

The absorption coefficient \eqref{eq:sigmaaexpr} expressed in wavenumber domain is now given by
\begin{equation}\label{eq:sigmaaexprnu}
\sigma_\mrm{a}(\nu)=\frac{\pi\eta_0}{3\hbar}\nu(1-\eu^{-\beta h\mrm{c}_0\nu})\frac{1}{\pi}\Re\left\{\widetilde{C}(\nu)\right\}, %I(\nu)
\end{equation}
where the SI-units of the factor $\pi\eta_0/3\hbar$ is given in $A^{-2}s^{-2}$, $\widetilde{C}(\nu)$ in $A^2s^2m^3$ and $\sigma_\mrm{a}$ in \unit{m^2} as desired.
Now, the line strength parameters $S_n^\prime$ of the HITRAN database are given in length dimension (typically in \unit{cm}) and are defined in
such a way that the absorption coefficient due to any specific transition is given by
$\sigma_{\mrm{a},n}(\nu)=S_n^\prime f_n(\nu)$ where $f_n(\nu)$ is a normalized line shape function such as the Lorentzian, Gaussian or Voigt, \cf \cite[Eq.~(8)--(10)]{HITRANdefs}.
Hence, by assuming that the slowly varying frequency dependency of the factor $\nu(1-\eu^{-\beta h\mrm{c}_0\nu})$ in \eqref{eq:sigmaaexprnu} above
is not involved in this definition and by writing $I_n(\nu)=S_nf_n(\nu)$ for a specific transition, it is found that
\begin{equation}
S_n^\prime=\frac{\pi\eta_0}{3\hbar}\nu_n(1-\eu^{-\beta h\mrm{c}_0\nu_n})S_n.
\end{equation}
The absorption coefficient \eqref{eq:sigmaaexprnu} based on the sum of isolated Lorentzian lines 
as in \eqref{eq:Cnulinesk0} with $a_n^2=S_n$ can now be expressed as
\begin{equation}\label{eq:sigmaaexprnu1}
\sigma_\mrm{a}(\nu)=\nu(1-\eu^{-\beta h\mrm{c}_0\nu})
\frac{1}{\pi}\Re\left\{\sum_n \frac{S_n^{\prime\prime}}{\gamma_n^\prime-\iu(\nu-\nu_n)}\right\},
\end{equation}
where
\begin{equation}\label{eq:Snbis}
S_n^{\prime\prime}=\frac{\pi\eta_0}{3\hbar}S_n=\frac{S_n^\prime}{\nu_n(1-\eu^{-\beta h\mrm{c}_0\nu_n})}.
\end{equation}
It is not difficult to see that the other spectral functions expressed above can be modified in the same way.
We have for the exact solution \eqref{eq:extactCnukeq0} and \eqref{eq:extactC1nukeq0}
\begin{equation}\label{eq:sigmaaexprnu2}
\sigma_\mrm{a}(\nu)=\nu(1-\eu^{-\beta h\mrm{c}_0\nu})
\frac{1}{\pi}\Re\left\{\widetilde{C}^{\prime\prime}(\nu)\right\},
\end{equation}
where 
\begin{equation}\label{eq:extactCnubis}
\widetilde{C}^{\prime\prime}(\nu)=\frac{\widetilde{C}_1^{\prime\prime}(\nu)}{1-\frac{v_\mrm{s}^\prime}{C^{\prime\prime}(0)}\widetilde{C}_1^{\prime\prime}(\nu)},
\end{equation}
and
\begin{equation}\label{eq:extactC1nubis}
\widetilde{C}_1^{\prime\prime}(\nu)
=\sum_n \frac{S_n^{\prime\prime}}{\gamma_{0n}^\prime+v_\mrm{s}^\prime\frac{S_n^{\prime\prime}}{C^{\prime\prime}(0)}-\iu\left(\nu-\nu_{0n}-\delta_{0n}^\prime\right)},
\end{equation}
and where $C^{\prime\prime}(0)=\sum_n S_n^{\prime\prime}$.

In the more general setting including Doppler shift we can use the expression \eqref{eq:sigmaaexprnu2}
with $\widetilde{C}^{\prime\prime}(\nu)=\frac{\pi\eta_0}{3\hbar}\widetilde{C}(\nu)$ and where \eqref{eq:Cnulines} becomes 
\begin{equation}\label{eq:Cnulinesbis}
\widetilde{C}^{\prime\prime}(\nu)
=\sum_n \frac{{a_n^{\prime\prime}}^2D_{n}(\nu)}{1-\beta^\prime D_n(\nu)},
\end{equation}
where ${a_n^{\prime\prime}}^2=\frac{\pi\eta_0}{3\hbar}a_n^2$. The first order Rosenkranz parameter
$a_n^{\prime\prime}$ is then finally given by \eqref{eq:anfirstorderSCnu}, which becomes
\begin{equation}\label{eq:anfirstorderSCnubis}
a_n^{\prime\prime}=M_n^{\prime\prime}+\frac{\iu p\widehat{v}_\mrm{s}^\prime M_n^{\prime\prime}}{C^{\prime\prime}(0)}\sum_{k\neq n}\frac{S_k^{\prime\prime} }{\nu_{0n}-\nu_{0k}},
\end{equation}
and where $M_n^{\prime\prime}=\sqrt{S_n^{\prime\prime}}$.
The line coupling parameter $\widehat{v}_\mrm{s}^\prime$ defined by \eqref{eq:vssol1nu} is also computed accordingly.

It should be noticed that the scaling \eqref{eq:Snbis} is only required
when we take the fluctuation-dissipation theorem into account by considering \eqref{eq:sigmaaexprnu}
and the factor $\nu(1-\eu^{-\beta h\mrm{c}_0\nu})$ has a significant frequency dependency over the band of interest.
This will be the case in the lower frequency ranges such as with the
second numerical example below dealing with the millimeter wave absorption of atmospheric water vapor and oxygen.
At higher frequencies, we can usually ignore the influence of the fluctuation-dissipation theorem
and consider the ratio $\nu(1-\eu^{-\beta h\mrm{c}_0\nu})/\nu_n(1-\eu^{-\beta h\mrm{c}_0\nu_n})\approx 1$.
This is then implemented above simply by putting $\sigma_\mrm{a}(\nu)=I(\nu)$ and replacing the line strengths $S_n$ in \unit{A^2s^2m^2}
for the HITRAN parameters $S_n^\prime$ which are usually given in length dimension \unit{cm}.

\section{Numerical examples}\label{sect:NumEx}

The capability of the modified projection method to model the narrowband as well as the broadband aspects of the absorption spectra of a gas
is illustrated below by using two different examples. The first example is with carbon dioxide in its ro-vibrational $\nu_3$-band at 2349\unit{cm^{-1}},
and the second is with moist air in the millimeter range up to 400\unit{GHz}.  Both examples are for typical low altitude tropospheric conditions.
All line calculations have been derived solely from the basic spectroscopic parameters: transition frequency, line strength and line width and shift, 
which have been retrieved from the HITRAN database \cite{Gordon+etal2022}.
The expressions \eqref{eq:sigmaaexprnu1} and \eqref{eq:sigmaaexprnu2} have been employed in all calculations, even though the fluctuation-dissipation theorem only has a very minor effect for the 
calculation of the ro-vibrational band of carbon dioxide. For the computation of the millimeter wave absorption of moist air, however, the use of \eqref{eq:sigmaaexprnu1} and \eqref{eq:sigmaaexprnu2}
are vital for achieving convergence with respect to the number of included lines. In all calculations below 
we have therefore included all the molecular transitions that are listed in the data base within the computational domain
of interest, plus all lines within 300\unit{cm^{-1}} outside to secure the convergence of the absorption coefficient.

\subsection{The $\mrm{CO}_2$ $\nu_3$-band}

\begin{figure}
\begin{center}
\includegraphics[width=0.48\textwidth]{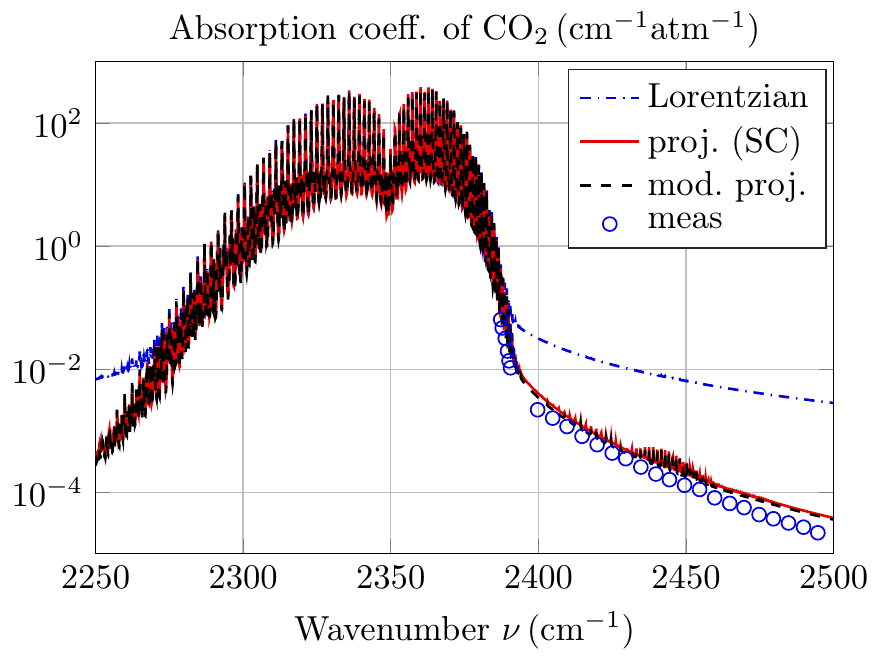}
\end{center}
\vspace{-5mm}
\caption{Relative absorption coefficient for the $\nu_3$ $\mrm{CO}_2$-band in dry air at $T=20$\unit{\degree C} and total pressure $p=1$\unit{atm}.
The blue dashdotted line (Lorentzian) indicates the sum of isolated Lorentzian lines, the red solid line (proj.~(SC)) the basic projection method 
and the black dashed line (mod.~proj.) the modified projection method.
The blue rings (meas) indicate the corresponding measurement data which have been visually interpreted from \cite[Fig.~2]{Tonkov+Filippov2003}. }
\label{fig:Projection_fig10}
\end{figure}

\begin{figure}
\begin{center}
\includegraphics[width=0.48\textwidth]{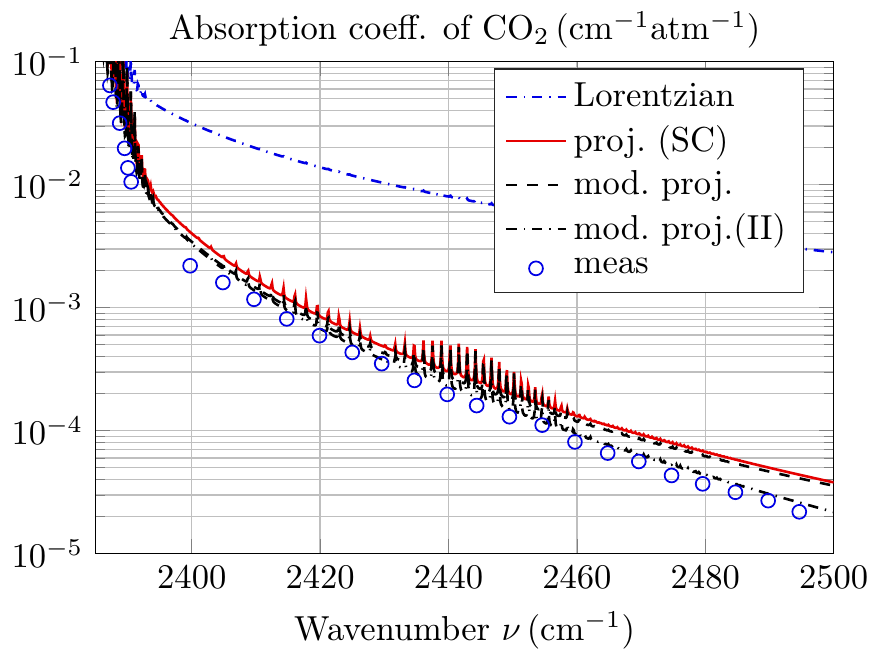}
\end{center}
\vspace{-5mm}
\caption{Same plot as in Fig.~\ref{fig:Projection_fig10} focusing on the far wing at 2385-2500\unit{cm^{-1}}.
Here, the black dashdotted line (mod. proj.(II)) corresponds to a small 0.5\unit{\%} increase of the  line coupling transfer rates where $\widehat{v}_\mrm{s}^\prime=1.005\cdot\widehat{v}_\mrm{s}^{\mrm{ls}\prime}$.}
\label{fig:Projection_fig11}
\end{figure}

\begin{figure}
\begin{center}
\includegraphics[width=0.48\textwidth]{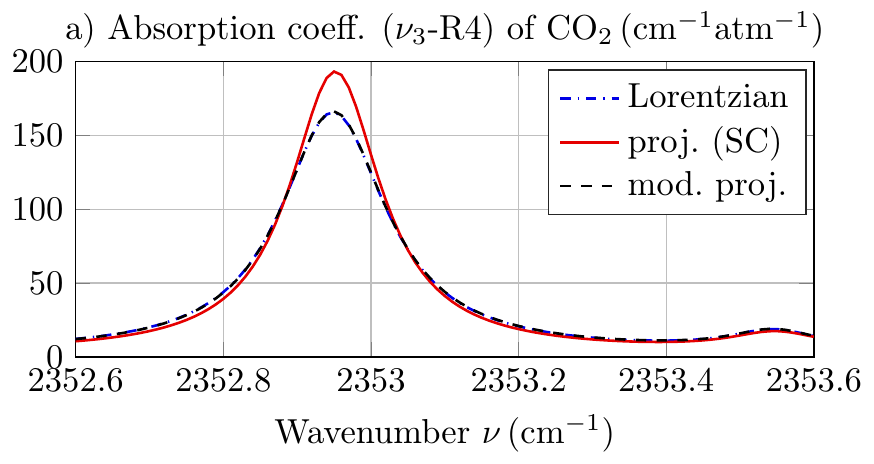}
\includegraphics[width=0.48\textwidth]{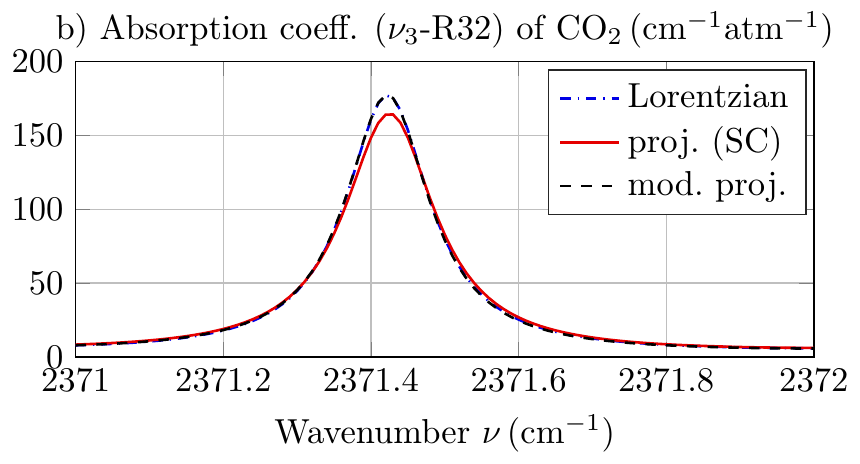}
\end{center}
\vspace{-5mm}
\caption{Same plot as in Fig.~\ref{fig:Projection_fig10} focusing on the rotational R4 and R32 lines of the $\nu_3$ R-branch, respectively.
Here, the modified projection (mod.~proj.) is virtually coincident with the adequate Lorentzian, and the basic SC method is in error.}
\label{fig:Projection_fig12o13}
\end{figure}

In Figs.~\ref{fig:Projection_fig10} through \ref{fig:Projection_fig12o13} are shown the relative absorption coefficient for the $\nu_3$ $\mrm{CO}_2$-band in dry air 
at $T=20$\unit{\degree C} and total pressure $p=1$\unit{atm}. %Gordon+etal2022
The input parameters $\gamma_{0n}$ and $\delta_{0n}$ are calculated for $1$\unit{\%} $\mrm{CO}_2$  and the absorption coefficient
is then scaled for path length in \unit{cm} and partial pressure in \unit{atm}.
In Figs. \ref{fig:Projection_fig10} and \ref{fig:Projection_fig11} are also included a comparison to measurement data which 
have been visually interpreted from \cite[Fig.~2]{Tonkov+Filippov2003} as indicated here with the blue rings.
The blue dashdotted lines indicate the sum of isolated Lorentzian lines, the red solid lines the basic projection (SC) method \eqref{eq:extactCnukeq0} together with 
\eqref{eq:vssol2nu} and \eqref{eq:extactC1nukeq0SC} and the black dashed lines the modified projection method \eqref{eq:extactCnukeq0}
together with  \eqref{eq:vssol1nu} and \eqref{eq:extactC1nukeq0} and where $v_\mrm{s}^\prime=p\widehat{v}_\mrm{s}^{\mrm{ls}\prime}$. 
In Fig.~\ref{fig:Projection_fig11} the parameter $\widehat{v}_\mrm{s}^\prime$ has furthermore been chosen as
$\widehat{v}_\mrm{s}^\prime=1.005\cdot\widehat{v}_\mrm{s}^{\mrm{ls}\prime}$ 
with the purpose to fine tune the absorption for a better match in the far wing and which is indicated here by the black dashdotted line.
The latter is a small modification of the line coupling transfer rates that does not affect the accuracy close to the line centers.
In Fig.~\ref{fig:Projection_fig12o13} is finally shown a close up view of two representative R-branch lines of the $\nu_3$-fundamental.
The small ``bump'' seen at 2353.55\unit{cm^{-1}} is due to the transition $01^10\rightarrow 01^11$ R23e.

It can now be observed in Figs.~\ref{fig:Projection_fig10} and \ref{fig:Projection_fig11} how the modified projection method is able to mimic and even improve the prediction 
of the basic projection (SC) method in the far wing region at the same time as the prediction of the modified method
is virtually coincident with the adequate Lorentzian 
close to the center of isolated lines where the basic SC method is in error, \cf  Fig.~\ref{fig:Projection_fig12o13}.
It should finally be noted that the basic projection (SC) method can not be successfully scaled in the same way as the modified projection method. 
In this example, it turns out that a similar scaling of the SC method by $\widehat{v}_\mrm{s}^\prime=0.6\cdot\widehat{v}_\mrm{s}^{\mrm{ls}\prime}$ (40 \% decrease
of $\widehat{v}_\mrm{s}^{\mrm{ls}\prime}$) will give the adequate far wing correction, 
but the line center resonances will then be completely destroyed. This is of course due to the fact that the scaling of the SC method
\eqref{eq:GammaSC} affects the diagonal elements as well as the line coupling transfer rates, whereas a similar scaling of the modified projection method \eqref{eq:GammaSCmod} 
does only affect the line coupling transfer rates and which therefore has a very minor effect on the line centers.

\subsection{The $\mrm{O}_2$-$\mrm{H}_2\mrm{O}$ millimeter-band}

\begin{figure}
\begin{center}
\includegraphics[width=0.48\textwidth]{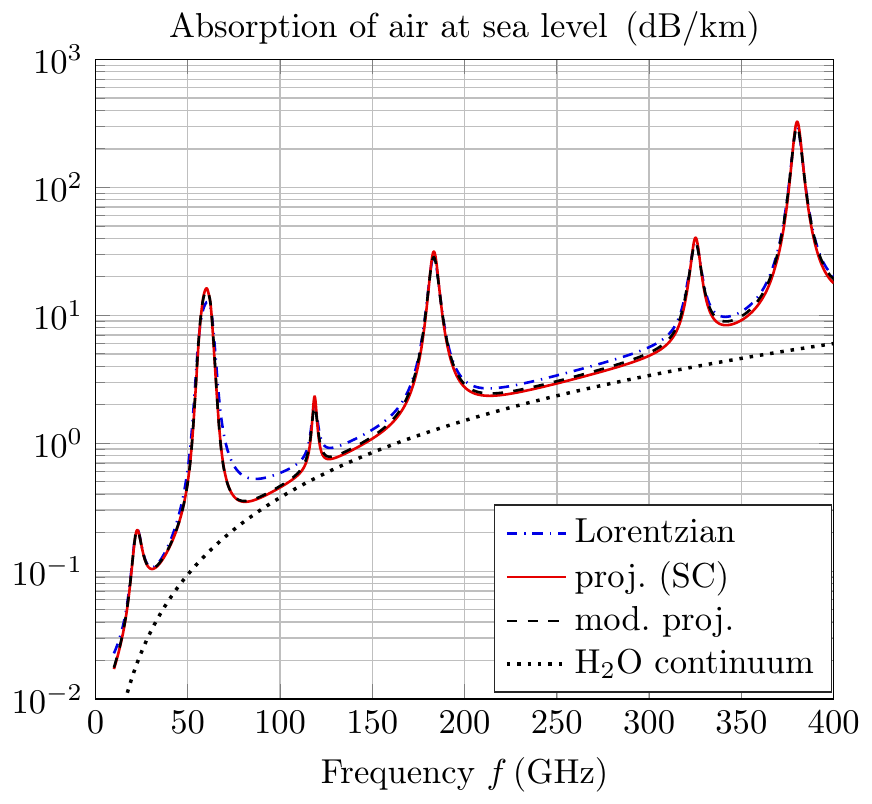}
\end{center}
\vspace{-5mm}
\caption{Absorption of moist air in the millimeter range at $T=15$\unit{\degree C}, total pressure $p=1$\unit{atm} and 60\unit{\%} humidity.
As before, the blue dashdotted line (Lorentzian) indicates the sum of isolated Lorentzian lines, the red solid line (proj.~(SC)) the basic projection method 
and the black dashed line (mod. proj.) the modified projection method. The contribution from the water vapor continuum absorption
is indicated by the black dotted line.}
\label{fig:Projection_fig6Ex1a} 
\end{figure}

In Fig.~\ref{fig:Projection_fig6Ex1a} is shown
the modeled millimeter range absorption of moist air at sea level where both the basic (proj.~(SC)) as well as the modified projection method (mod.~proj.) indicate results which are seemingly almost identical with
the predictions made in \cite[Fig.~4-6a on p.~124]{Richards+etal2010}. Here, the temperature is $T=15$\unit{\degree C} (288\unit{K}), 
the total pressure is $1$\unit{atm} and the radiatively active molecules consist of 21\unit{\%} oxygen and a water vapor content corresponding to 60\unit{\%} humidity. 
In this computation we have also included the water vapor continuum absorption (the black dotted line) calculated as a combination of the MPM87 and MPM93 
empirical models as proposed by Rosenkranz \cite{Rosenkranz1998}. 
Here, the total absorption coefficient is given by $\alpha=\alpha_\mrm{line}+\alpha_\mrm{cont}$ where $\alpha_\mrm{line}=N\sigma_\mrm{a}$ is the
usual line contribution of the radiator based on expressions like \eqref{eq:sigmaaexprnu1} and \eqref{eq:sigmaaexprnu2}, 
$N$ its number density and where the continuum contribution is given by the following expression 
\begin{equation}
\alpha_\mrm{cont}=f^2\left(\frac{300}{T}\right)^3\left(C_\mrm{f}p_\mrm{f}p_\mrm{s}+C_\mrm{s}p_\mrm{s}^2\right)
\end{equation}
where $f$ is frequency in \unit{GHz}, $T$ temperature in \unit{K} and $p_\mrm{f}$ and $p_\mrm{s}$ the partial pressure of air and water vapor, respectively, both in \unit{kPa}.
The coefficient for foreign (air) broadening is $C_\mrm{f}=2.38\cdot 10^{-7}$ (MPM87) and for self (water vapor) broadening
 $C_\mrm{s}=7.8\cdot 10^{-6}\left(\frac{300}{T}\right)^{4.5}$ (MPM93), both in \unit{dB/km/GHz^2/kPa^2}.
 
\begin{figure}
\begin{center}
\includegraphics[width=0.48\textwidth]{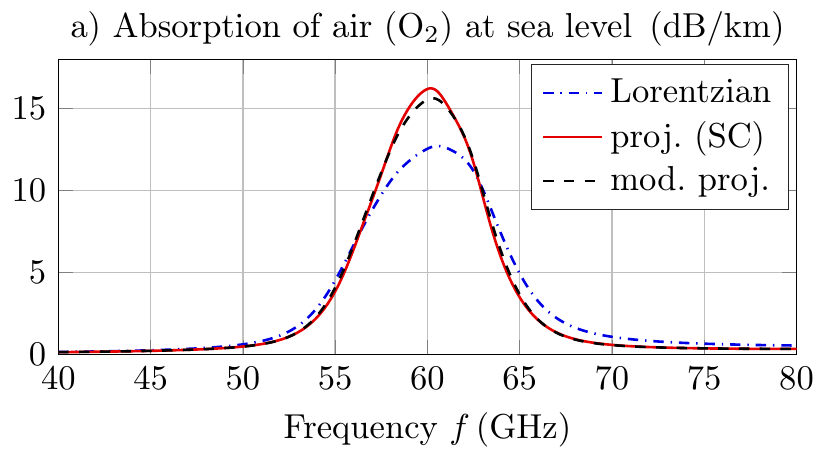}
\includegraphics[width=0.48\textwidth]{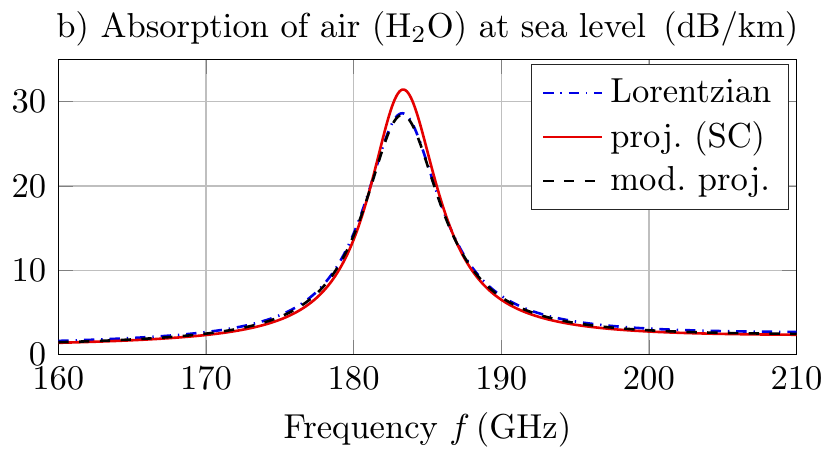}
\end{center}
\vspace{-5mm}
\caption{Same plot as in Fig.~\ref{fig:Projection_fig6Ex1a} focusing on a) the oxygen 60\unit{GHz} band and b) the water vapor 183\unit{GHz} band, respectively.
Notice that the modified projection method follows the basic projection (SC) method in the oxygen band where line mixing is important, but it follows instead the Lorentzian 
close to the dominating isolated water vapor line at 183.3\unit{GHz} where the SC method is largely deviating.}
\label{fig:Projection_fig6Ex1boc}
\end{figure}

 In Fig.~\ref{fig:Projection_fig6Ex1a}, we can now observe that the Lorentzian model is giving excess absorption, 
 and in particular in between the two resonances of oxygen at 60\unit{GHz} and 120\unit{GHz}, respectively,
 as well as in the upper wing of the latter. This discrepancy is alleviated by both of the present line mixing methods.
 In Figs~\ref{fig:Projection_fig6Ex1boc} a) and b) are shown the corresponding close up views of the oxygen 60\unit{GHz} resonance
 and the water vapor resonance at 183.3\unit{GHz}, respectively. 
 It may be noticed here that the corresponding experimental value of the 60\unit{GHz} peak absorption of 
dry air at 1\unit{atm} and 15\unit{\degree C} is very close to 15\unit{dB/km}, \cf \cite[Fig.~2]{Makarov+etal2011}.
 As can be seen in these plots, the modified projection method follows
 closely the basic projection method (SC) in the oxygen band at 60\unit{GHz} where lines overlap and line mixing is adequate.
 However, in the water vapor band at 183.3\unit{GHz} the modified projection method follows instead tightly the
 Lorentzian where the dominating rotational line of water is virtually isolated and the basic SC method is in error.
 
\begin{figure}
\begin{center}
\includegraphics[width=0.48\textwidth]{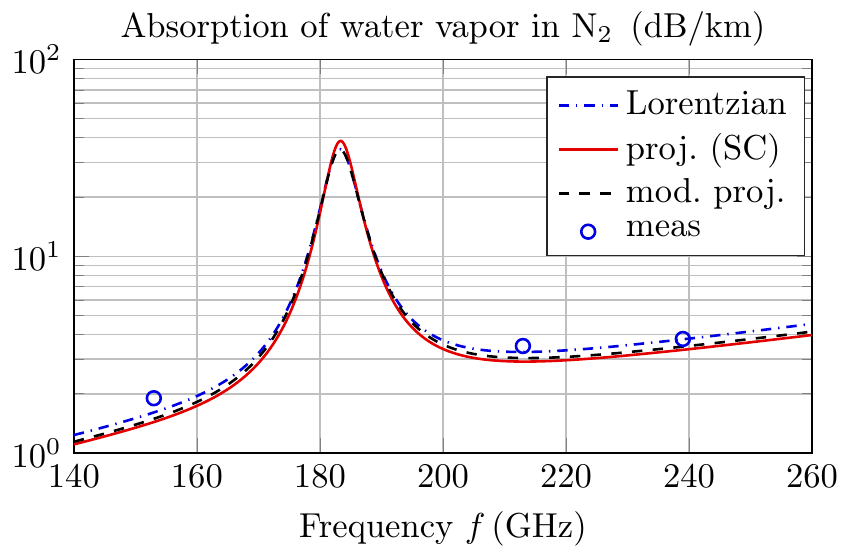}
\end{center}
\vspace{-5mm}
\caption{Absorption of an $\mrm{N}_2$-$\mrm{H}_2\mrm{O}$ mixture at $T=23$\unit{\degree C}, total pressure $1$\unit{atm} and water vapor pressure $p_\mrm{s}=1/76$\unit{atm} (10 torr).  
The blue rings (meas) indicate measurements according to \cite[Fig.~10 on p.~420]{Bauer+etal1995}.}
\label{fig:Projection_fig6Ex2}
\end{figure}

In Fig.~\ref{fig:Projection_fig6Ex2} is finally shown the modeling results for the 183.3\unit{GHz}
absorption band of an $\mrm{N}_2$-$\mrm{H}_2\mrm{O}$ mixture at $T=23$\unit{\degree C} (296\unit{K}),
total pressure $1$\unit{atm} and water vapor pressure $p_\mrm{s}=1/76$\unit{atm} (10 torr). 
The modeling results are compared to measured data according to \cite[Fig.~10 on p.~420]{Bauer+etal1995}.
As we can see here, the measured values are in fact closest to the Lorentzian model. 
We should remember, however, that this is in the wings of an isolated line.
And it makes sense since the MPM87/93 parameters used above have been calibrated for the sum of Lorentzian lines, \cf \cite[Eq. (4)]{Rosenkranz1998},
and line mixing typically reduces the absorption in the wings.
However, the discrepancy is not very large and the MPM87/93 water vapor continuum model together with the proposed
modified projection method appears to provide a useful model for millimeter waves taking line mixing effects into account, 
as illustrated in Figs.~\ref{fig:Projection_fig6Ex1a} through \ref{fig:Projection_fig6Ex2}.

\section{Summary}\label{sect:Summary}
A modified projection approach to line mixing has been presented which is based on a simple adjustment 
of the strong collision (SC) method introduced by Bulanin, Dokuchaev, Tonkov and Filippov. It has been demonstrated how basically any 
desired isolated line model encompassing uncorrelated collisions can be used as diagonal elements of the collisional relaxation matrix, 
at the same time as the SC line coupling transfer rates can be retained and optimally scaled to provide a proper
far wing behavior. The method thus provides a simple and flexible compromise towards a unified treatment for all species in all parts of the spectrum.

In particular, by following the ideas of \cite{Tonkov+etal1996,Tonkov+Filippov2003}, it is possible to express explicitly an exact solution to the line mixing problem 
in the case of uncorrelated collisions, pure pressure broadening and velocity independent broadening and shifts parameters.
To include Doppler broadening one can readily apply the first order Rosenkranz approximation. Notably, within the present context 
with uncorrelated collisions and line independent Dicke narrowing, the Rosenkranz approximation will also allow the diagonal elements of the relaxation matrix to depend on speed 
and where analytical results exist with the quadratically speed dependent parameters associated with the Hartmann-Tran (HT) profile \cite{Ngo+etal2013}.

The method has been illustrated by using numerical examples including comparisons to published measured data 
in two specific cases concerning the absorption of carbon dioxide in its infrared $\nu_3$ ro-vibrational band at 2349\unit{cm^{-1}}, as well as 
of atmospheric water vapor and oxygen in relevant millimeter bands up to 400\unit{GHz}.

\appendix
\subsection{Integrals involving the Faddeeva function}\label{sect:integrals}

Some of the important integral relationships involving the Maxwell-Boltzmann distribution as well as the Faddeeva function used in this paper are summarized below.
The Faddeeva function is an entire function defined by 
\begin{equation}
w(z)=\eu^{-z^2}\mrm{erfc}(-\iu z)
\end{equation}
where $\mrm{erfc}(z)$ is the complementary error function
defined by $\mrm{erfc}(z)=\frac{2}{\sqrt{\pi}}\int_z^\infty\eu^{-t^2}\mrm{d}t$, see \eg \cite[Eq.~(7.2.1) -- (7.2.3)]{Olver+etal2010}. 
The Faddeeva function is of fundamental importance and very convenient to use in spectroscopic modeling due to the vast literature 
that is available on the theory, algorithms and computer codes for its efficient numerical evaluation, see \eg \cite{Armstrong1967,Schreier1992,Schreier2011,Abrarov+Quine2011,Tennyson+etal2014}.
The Maxwell-Boltzmann velocity distribution is given by $f(\bm{v})=(\sqrt{\pi}\widetilde{v})^{-3}\eu^{-(v/\widetilde{v})^2}$
where $v=|\bm{v}|$, $\widetilde{v}=\sqrt{2k_\mrm{B}T/\mu_\mrm{r}}$ is the most probable speed, $k_\mrm{B}$ the Boltzmanns constant, 
$T$ the temperature and $\mu_\mrm{r}$ the mass of the radiator, \cf \cite[Chapt.~5]{Blundell+Blundell2010}.

The first integral of interest here is the standard result
\begin{equation}\label{eq:CtDopplerresult}
\int_{\bm{v}}f(\bm{v})\eu^{-\iu\bm{k}\cdot\bm{v} t}\mrm{d}\bm{v}=\eu^{-(k\widetilde{v}t/2)^2},
\end{equation}
where $\bm{k}=k\hat{\bm{k}}$ is the wave vector associated with the incident radiation, see \eg \cite{Rautian+Sobelman1967}. 
The integral identity \eqref{eq:CtDopplerresult} can readily be obtained by a substitution to spherical coordinates and is valid for all $t\in\R$.

By completing the squares in the exponent and employing the definition of the complementary error function, 
one can also derive the following Fourier-Laplace transform
\begin{equation}\label{eq:ComegaDoppler5}
\int_0^\infty\eu^{-(k\widetilde{v}t/2)^2}\eu^{\iu\omega t}\mrm{d}t
=\frac{\sqrt{\pi}}{k\widetilde{v}}w\left(\frac{\omega}{k\widetilde{v}}\right),
\end{equation}
which is valid for all $\omega\in\C$. Another useful integral identity can also be derived by inserting the result \eqref{eq:CtDopplerresult}
into \eqref{eq:ComegaDoppler5} and changing the order of integration to yield
\begin{equation}\label{eq:ComegaDoppler6}
\int_{\bm{v}} \frac{f(\bm{v})\mrm{d}\bm{v}}{\iu\bm{k}\cdot\bm{v}-\iu\omega}
=\frac{\sqrt{\pi}}{k\widetilde{v}}w\left(\frac{\omega}{k\widetilde{v}}\right),
\end{equation}
but which is now valid only for $\Im\{\omega\}>0$. The latter restriction is usually not a problem since the resulting right-hand side can be analytically extended to the whole complex plane. 
An important example is the relation 
\begin{equation}\frac{1}{\pi}\Re\left\{\frac{\sqrt{\pi}}{k\widetilde{v}} w\left(\frac{\omega}{k\widetilde{v}}\right)\right\}
=\frac{1}{\sqrt{\pi}k\widetilde{v}}\eu^{-\omega^2/k^2\widetilde{v}^2}
\end{equation}
where $\omega\in\R$, and which is a well known formula for Doppler broadening, \cf \eg \cite{Rautian+Sobelman1967,Hartmann+etal2021,Tennyson+etal2014}.
The integral identities \eqref{eq:CtDopplerresult}, \eqref{eq:ComegaDoppler5} and \eqref{eq:ComegaDoppler6}
are standard integrals which are employed in many papers, such as \eg \cite{Ngo+etal2013,Ciurylo+Pine2000}.

%\bibliographystyle{IEEEtran}
%\bibliography{total}


\end{document}